\documentclass[colorlinks]{goose-article}

\usepackage{ragged2e}
\usepackage{ltablex}
\usepackage[T1]{fontenc}

\title{Thermal origin of quasi-localised excitations in glasses}

\author[1]{Wencheng~Ji}
\author[1]{Tom~W.J.~de~Geus}
\author[1]{Marko~Popovi\'{c}}
\author[1]{Elisabeth~Agoritsas}
\author[1]{Matthieu~Wyart}

\affil[1]{%
    Institute of Physics,
    \'Ecole Polytechnique F\'ed\'erale de Lausanne (EPFL),
    CH-1015 Lausanne,
    Switzerland}

\contact{}
\header{Ji et al. (2020), \emph{Phys.\ Rev.\ E}, 102:062110, \doi{10.1103/PhysRevE.102.062110}, \arxivid{1912.10537}}
\hypersetup{pdfauthor={T.W.J. de Geus}}

\begin{document}

\twocolumn[
\begin{@twocolumnfalse}

\maketitle

\begin{abstract}
Key aspects of glasses are controlled by the presence of excitations in which
a group of particles can rearrange.
Surprisingly, recent observations indicate that their density is dramatically reduced
and their size decreases as the temperature of the supercooled liquid is lowered.
Some theories predict these excitations to cause a gap in the spectrum of
quasi-localised modes of the Hessian that grows upon cooling, while others predict a pseudo-gap ${D_L(\omega)} \sim \omega^\alpha$.
To unify these views and observations, we generate glassy configurations of controlled gap
magnitude $\omega_c$ at temperature ${T=0}$,
using so-called `breathing' particles,
and study how such gapped states respond to thermal fluctuations.
We find that
\textit{(i)}~the gap always fills up at finite $T$ with
${D_L(\omega) \approx A_4(T) \, \omega^4}$ and ${A_4 \sim \exp(-E_a / T)}$ at low $T$,
\textit{(ii)}~$E_a$ rapidly grows with $\omega_c$, in reasonable agreement with a simple
scaling prediction ${E_a\sim \omega_c^4}$ and
\textit{(iii)}~at larger $\omega_c$ excitations involve fewer particles, as we rationalise,
and eventually become string-like.
We propose an interpretation of mean-field theories of the glass transition,
in which the modes beyond the gap act as an excitation reservoir,
from which a pseudo-gap distribution is populated with its magnitude rapidly decreasing at lower $T$.
We discuss how this picture unifies the rarefaction as well as the decreasing size
of excitations upon cooling,
together with a string-like relaxation occurring near the glass transition.
\end{abstract}

\end{@twocolumnfalse}
]
\sloppy


\section{Introduction}
\label{sec:intro}

A key feature of structural glasses is that groups of particles can rearrange locally
between two metastable states.
This motion can be triggered by quantum or thermal fluctuations,
or mechanically by exerting an external stress or strain.
Such rearrangements are associated with different excitations.
At low temperature the dominant source of excitations are two-level systems (TLS) that stem from quantum tunnelling between the metastable states \cite{Phillips72,Anderson72,Phillips87}.
At higher temperatures, relaxation in supercooled liquids near the glass transition occurs via thermally activated events, observed to become more and more string-like upon cooling \cite{Donati98,Yu17}. Upon mechanical loading, at any temperature below the glass transition,
plasticity occurs when a group of particles becomes unstable.
In the potential energy landscape, this corresponds to a saddle-node bifurcation \cite{Maloney04,Heuer08}
and leads to a rearrangement denoted `shear transformation' \cite{Argon79}.
Understanding how temperature or system preparation controls the density of these excitations remains a challenge.
It is, however, a question of practical importance since: \textit{(i)}~the density of shear transformations
controls for instance the glass brittleness \cite{Vandembroucq2011,Ozawa18,Popovic18}, \textit{(ii)}~the rarefaction of activated events near the glass transition controls its
fragility \cite{Ediger96},
and \textit{(iii)}~the density of TLS (recently observed to be almost absent in
ultra-stable glasses \cite{Queen13,Perez14})
affects the decoherence in qubits \cite{Martinis05} important for quantum computing.
Finally, the possible unification of these excitations
into a common description is a fundamental problem for a prospective theory of glasses.

These localised excitations should affect the low-frequency spectrum of the Hessian of the energy landscape,
since groups of particles that can easily rearrange tend
to have a small linear restoring force \cite{Parshin07,Maloney04,Wencheng19}.
Indeed, in numerical glasses, such quasi-localised modes are found at
low-frequencies \cite{Schober93}.
Recently there has been a considerable effort to analyse them
\cite{Baity15,Lerner16,Mizuno17,Lerner18,Shimada18,Scalliet19,Wang19}.
In most glasses, it is found that in inherent structures\footnote{%
    Obtained by rapidly quenching an equilibrated liquid
    (at initial temperature $T$) to zero temperature.
},
the density of quasi-localised modes ${D_L (\omega) \approx A_4 (T) \, \omega^4}$,
with $\omega$ the frequency.
Most remarkably, ${A_4 (T)}$ is reduced by several decades as $T$
is reduced by ${30\%}$ \cite{Wang19,Rainone20}
(a similar finding was obtained for the density of TLS \cite{Khomenko19}).
Furthermore, quasi-localised modes also display a lower participation ratio at lower $T$.
A unifying explanation for these facts is currently missing\footnote{%
    TLS were proposed to be controlled by the so-called mosaic length scale that diverges
    at the Kauzmann temperature $T_K$ in mean field approaches \cite{Lubchenko01}.
    Yet this description predicts a growing (instead of decreasing) length scale
    and a mild (a factor 10 at most) decrease of density of excitations upon cooling.}.
On the theoretical side, two distinct approaches have been proposed.
On the one hand, the $\omega^4$ power law has been rationalised by making specific assumptions
on the disorder and by assuming modes as non-interacting \cite{Gurevich03,Gurarie03},
or by modelling a quench from ${T = \infty}$ and including interactions \cite{Wencheng19}.
On the other hand, in mean-field calculations in infinite dimensions for
temperatures below the mode-coupling temperature $T_c$~\cite{Cavagna09}
the spectrum of the Hessian becomes gapped
(excluding obvious long wave-length Goldstone modes that are always present).
Below $T_c$, the gap is predicted to grow as $T$ decreases
\cite{Lubchenko07,Biroli12,Parisi03}.
A gap was also predicted from real-space stability arguments in finite dimensions
for continuously polydisperse particles, at very low energies and
zero temperature \cite{Kapteijns19}.
Nevertheless, it is currently unclear if a gap truly exists in
finite dimension and at finite temperature.

\begin{figure}[ht]
    \centering
    \includegraphics[width=\linewidth]{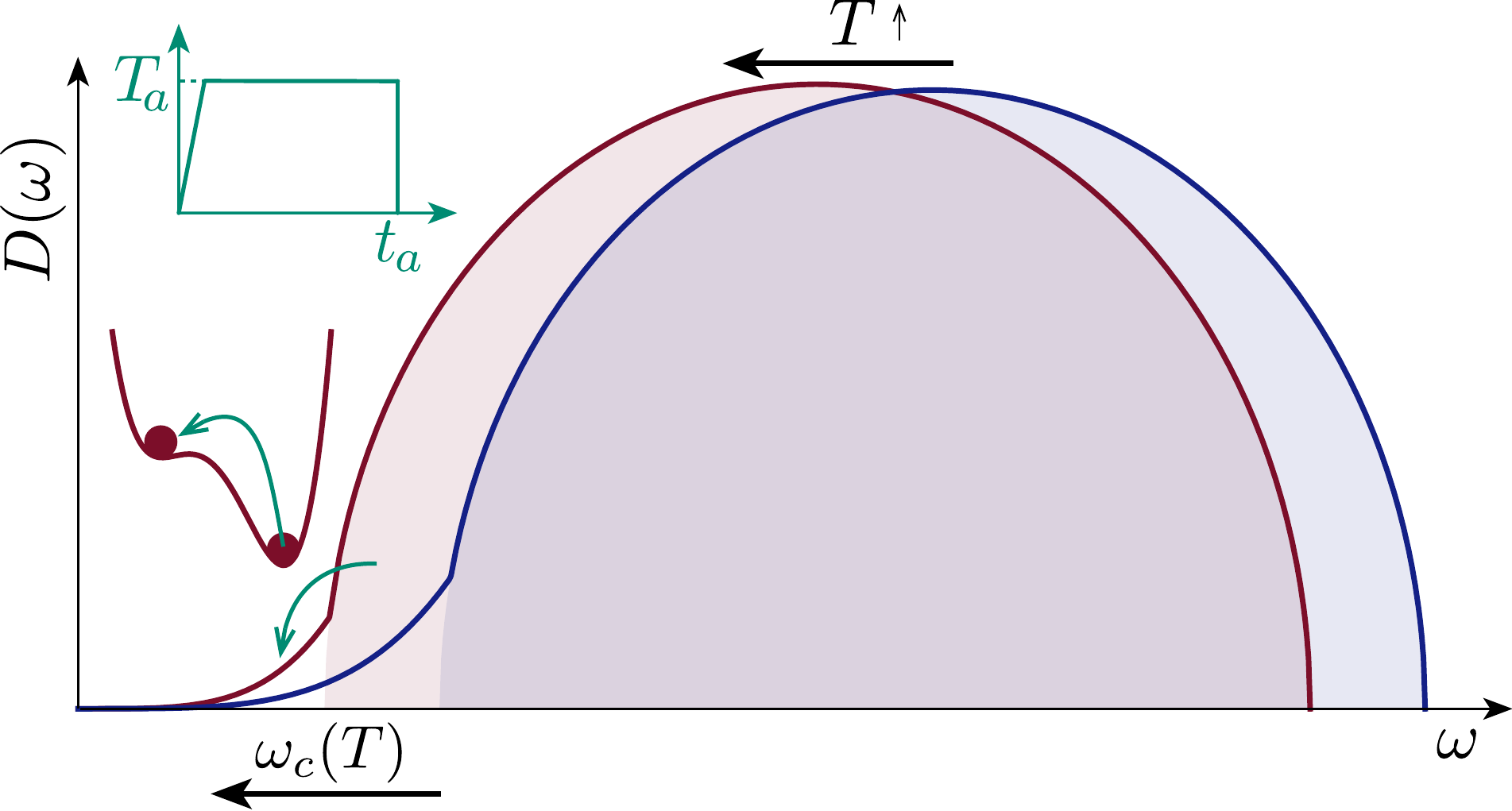}
    \caption{
    Schematic density of states for an equilibrated liquid at temperature $T$.
    When a gapped glass is heated to a temperature $T_a$ for a duration $t_a$,
    as sketched in the inset,
    modes beyond the gap act as a reservoir of excitations that can be thermally activated.
    It fills up the gap, leading, for small $\omega$, to a pseudo-gap ${D(\omega) \approx D_L(\omega)\approx A_4\; \omega^4}$.
    This effect is exponentially diminished if $\omega_c$ increases (corresponding to a decrease of $T$ as predicted by the infinite-dimensional mean-field description near the glass transition).
    }
    \label{fig:intro}
\end{figure}

In this article we seek a unifying scenario for these facts and different approaches,
by studying the stability of gapped spectra with respect to thermal fluctuations.
Specifically, we use `breathing' particles\footnote{
    `Breathing' particles is an alternative version of swap algorithms in which particles of
    different radii are exchanged \cite{Ninarello17}.}
\cite{Brito18}
in order to generate athermal ultra-stable glasses of controlled gap magnitude $\omega_c$.
Then, as sketched in \cref{fig:intro}, we transiently reheat these glasses,
with a standard molecular dynamics simulation, at a low
temperature $T_a$ for a duration $t_a$,
before quenching them back to zero temperature.
Our central results are that
\textit{(a)}~thermal fluctuations, even small, destroy the gap and we
recover a density ${D_L(\omega) \approx A_4 \, \omega^4}$;
the prefactor $A_4 (T_a, t_a)$ depends very mildly on $t_a$ but presents
an Arrhenius dependence on temperature
with ${A_4 \sim \exp(-E_a / T_a)}$ (in our temperature units the
Boltzmann constant ${k_B = 1}$).
\textit{(b)}~The activation energy $E_a$ rapidly increases with the gap magnitude $\omega_c$.
\textit{(c)}~We introduce a novel algorithm to decompose the
rearrangements into elementary excitations,
and find that they involve fewer particles for larger gap values,
and eventually become string-like for our largest gap.
We propose a scaling argument for their decreasing size.
Overall, these results suggest to describe equilibrated liquids perturbatively
as \emph{gapped states decorated by thermally activated excitations whose characteristic energy is
controlled by the gap itself},
leading to a contribution with ${A_4\sim \exp(-E_a(\omega_c(T))/T)}$.
We discuss the implications of this picture, sketched in \cref{fig:intro},
for the density of these various excitations, for their effect on plasticity and on low-temperature properties of glasses as well as for the glass transition.

\section{Generating gapped glasses}

To generate ultra-stable glasses displaying a finite gap,
we follow a procedure similar to \cite{Kapteijns19}.
We consider `breathing' particles whose individual size can vary according to an energetic cost
of characteristic stiffness $K$ (see \cref{sec:md}).
The particles interact with a repulsive potential, up to a finite cutoff radius,
chosen such that the potential remains continuous up to its third derivative \cite{Lerner17}
and thus allowing for a well-defined Hessian.
At a given temperature, this system is known to be thermodynamically equivalent
to a system of given (and continuous) polydispersity,
and can be simulated  using a usual molecular dynamics (MD).
Including this breathing
degrees of freedom leads to a giant shortening
of the equilibration time, comparable to that of swap algorithms \cite{Brito18,Berthier19b}.
In practice, we perform MD with breathing particles
for a long duration $t_p$ at a temperature $T_p(K)$,
chosen such as to minimise the energy of the states eventually obtained (see \cref{sec:sample}),
before quenching using a `FIRE' algorithm~\cite{Bitzek06} in which particles can still breathe.

\begin{figure}[ht]
    \centering
    \includegraphics[width=\linewidth]{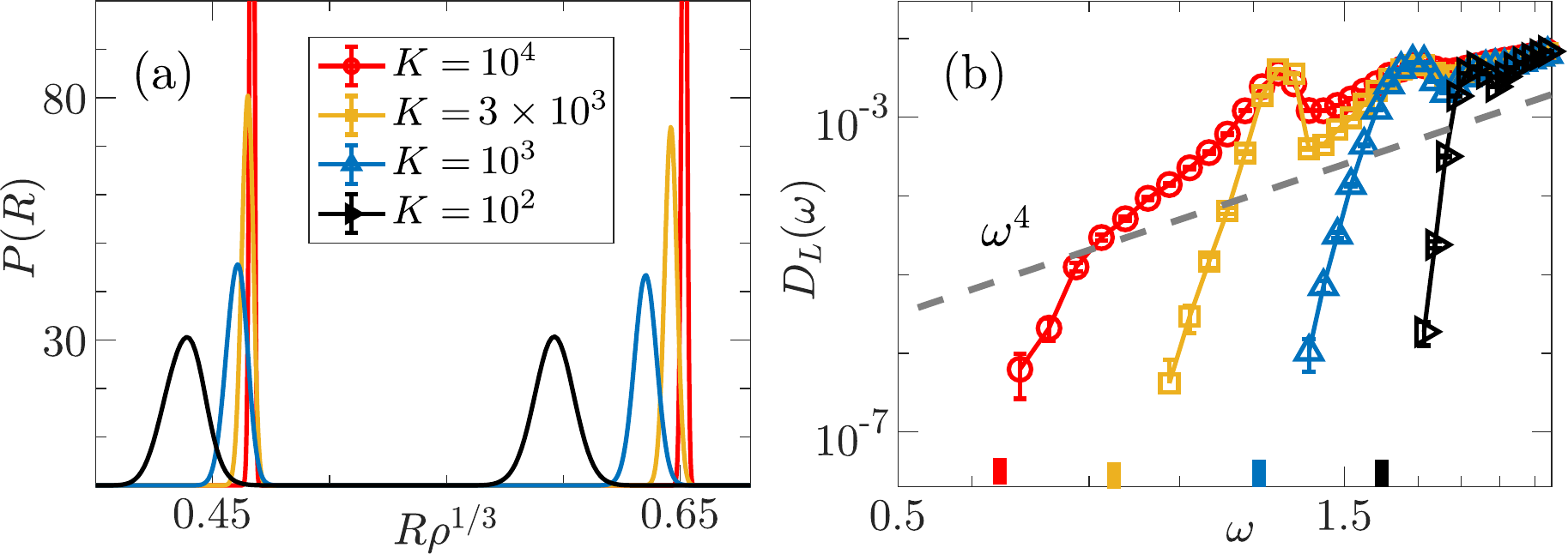}
    \caption{
        (a)~Distribution of particle radii, normalised by the number density
        ${\rho=N/\langle V \rangle}$, for different values of stiffness $K$
        (the particle sizes are narrowly distributed when $K$ is large).
        (b)~Density of quasi-localised modes displaying a finite gap~$\omega_c$,
        in contrast to the usual pseudo-gap scaling ${D_L(\omega) \sim \omega^4}$
        indicated with a dashed line.
        The gap values ${\omega_c \approx \{1.64,1.19,0.85,0.65\}}$ corresponding respectively to
        ${K=\{10^2, 10^3, 3 \times 10^3, 10^4\}}$ are indicated
        using ticks, following the same color code.
        Physically, decreasing~$K$ results to a larger gap and thus a more stable glass,
        and is associated to a larger polydispersity.
    }
    \label{fig:omega_c}
\end{figure}

The polydispersity obtained for various values of stiffness $K$ is shown in \cref{fig:omega_c}(a)
for ${N = 8000}$ particles, in three dimensions and at fixed pressure.
Next, we freeze the radius of each particle, and compute the usual Hessian of the potential energy:
its eigenvectors correspond to the vibrational modes of the glass,
and its eigenvalues are denoted $\omega^2$
since they correspond directly to the frequencies of vibrational modes,
as we take the particle mass to be unity.
Showing that these states are gapped requires considerable statistics;
in fact, we collect the spectra of ${n = 4000}$ independent realisations
(see \cref{sec:nomenclature} for a precise statement) and average them
in order to obtain the density of vibrational modes ${D(\omega)}$.
We emphasise that for the considered small system size, quasi-localised modes are already found below the first plane waves \cite{Lerner16}.

${D(\omega)}$ turns out to display a gap:
there are no quasi-localised modes below a finite frequency $\omega_c$.
Since we find $\omega_c$ to be even higher than the frequencies of the first
plane waves for ${K = \{ 10^2, 10^3 \}}$ we manually remove them in order to measure the density of
\emph{quasi-localised} modes ${D_L(\omega)}$,
as shown in \cref{fig:omega_c}(b).
We extract $\omega_c$ by fitting a power law ${D_L(\omega)\sim (\omega-\omega_c)^\zeta}$,
and obtain the values $\omega_c = \{ 1.64, 1.19, 0.85, 0.65 \}$,
for $K=\{10^2, 10^3, 3 \times 10^3, 10^4\}$, indicated
with markers in \cref{fig:omega_c}(b).
Note that if we consider instead the minimal frequency observed as an estimate for $\omega_c$,
our conclusions below are not affected (see \cref{sec:qlm}).

\section{Filling up the gap via thermal activation}

\begin{figure}[ht]
    \centering
    \includegraphics[width=\linewidth]{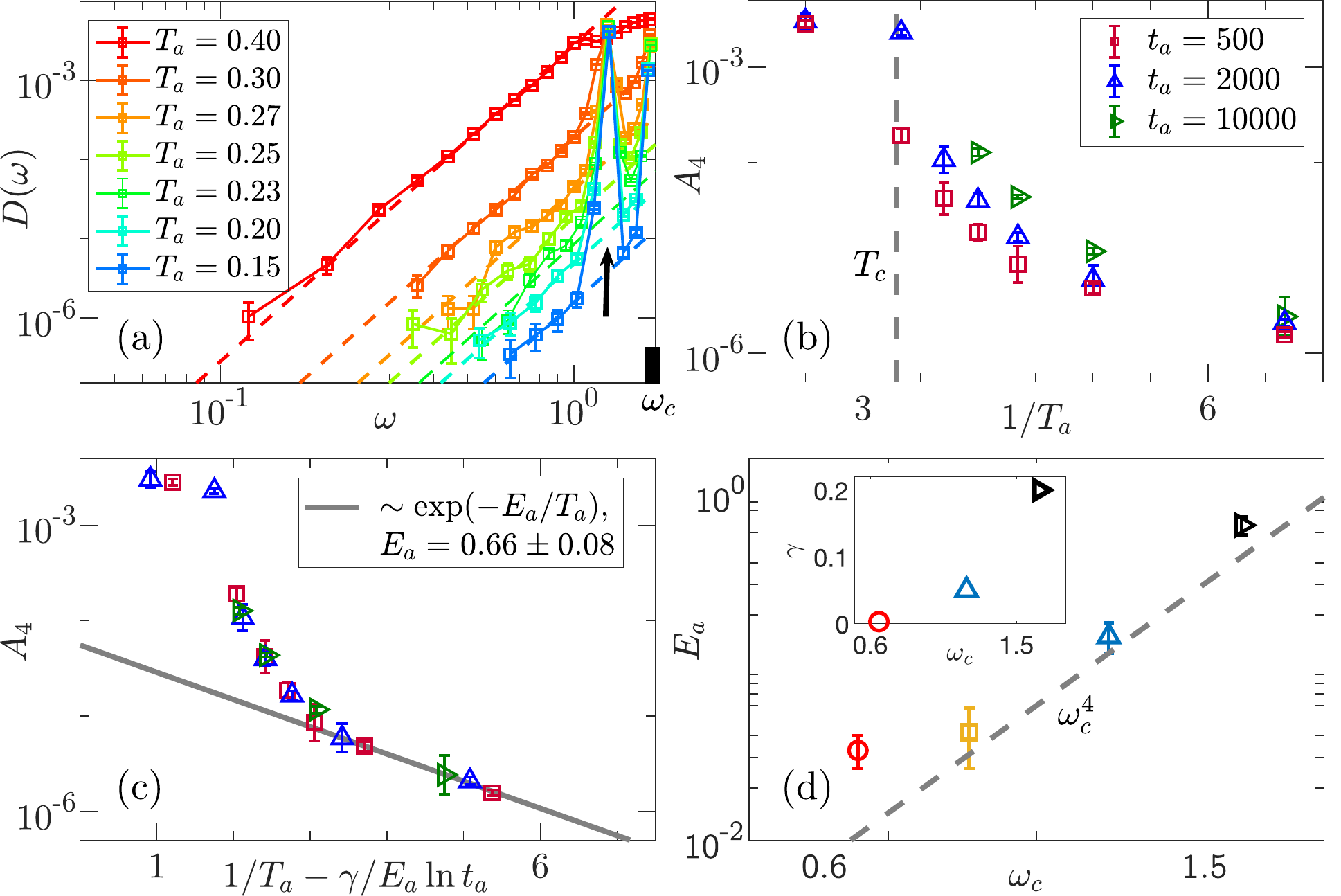}
    \caption{
        (a)~Density of soft modes $D(\omega)$ \emph{after} reheating for
        a fixed duration ${t_a = 500}$ at different temperatures $T_a$
        (following the protocol sketched in \cref{fig:intro}),
        at fixed gap $\omega_c = 1.64$ (the largest we generate, cf.\ \cref{fig:omega_c}).
        Note that the order of the legend matches the order of the curves.
        Furthermore, note that $D_L(\omega) = D(\omega)$ for $\omega < \omega_e$, with $\omega_e$ the frequency of the first plane wave, indicated with a black arrow.
        For reference, the mode-coupling temperature in this system is ${T_c \approx 0.3}$, and the gap, ${\omega_c = 1.64}$, is indicated by a black tick.
        We emphasise that, before reheating, we had ${D_L(\omega < \omega_c)=0}$,
        so that the corresponding modes have been activated by thermal fluctuations.
        (b)~Prefactor $A_4$ as a function of reheating temperature $T_a$
        for different durations $t_a$.
        (c)~Collapse of the different curves ${A_4(T_a,t_a)}$,
        supporting the functional form ${A_4 = f(t_a^\gamma \exp(-E_a / T_a))}$ where
        the function $f$ is
        linear at small argument,
        indicating an Arrhenius behaviour at small $T_a$.
        (d)~Typical energy scale $E_a$ \emph{vs} the initial gap $\omega_c$,
        together with the associated scaling prediction ${E_a \sim \omega_c^4}$ (dashed line).
        \textit{Inset:}~Dynamical exponent $\gamma$ as a function of $\omega_c$.
    }
    \label{fig:reheat}
\end{figure}

To test the robustness of gapped states to thermal fluctuations,
we reheat our samples to a temperature $T_a$ and run MD simulations for a duration $t_a$,
before applying an instantaneous quench to zero temperature.
This procedure is sketched in \cref{fig:intro} (and further detailed in \cref{sec:md}),
and is entirely performed at fixed particle radii.
Upon reheating, local rearrangements\footnote{
    We define a rearrangement by a finite norm of the displacement field that results from reheating,
    see \cref{sec:nomenclature} for details.}
are thermally triggered
(though less than 50\% of the samples do rearrange at the lowest temperature that we probe\footnote{For our system size, at $t_a = 500$ (the time scale we use later), the temperature for which we have in average one rearrangement per realisation $T_a^*(\omega_c=1.64)\approx0.17$, which we estimate using the fit of the Arrhenius-like behaviour below.}),
consequently modifying the spectrum.

In \cref{fig:reheat}(a) the low-frequency tail of $D(\omega)$
is shown for our stablest system (with ${\omega_c = 1.64}$) for $t_a=500$ and varying $T_a$.
Note that the acquisition of sufficient statistics required about $10^5$ CPU hours.
We always find that the gap is replaced by a pseudo-gap, compatible with the standard scaling:
\begin{equation}
    \label{eq:Dw}
    D_L(\omega) \approx A_4(T_a,t_a) \, \omega^4 .
\end{equation}
The prefactor $A_4$ characterises the density of quasi-localised excitations,
and is extracted by fitting \cref{eq:Dw} for ${\omega < \omega_e}$,
where $\omega_e$ is the frequency of the first plane wave (see \cref{fig:omega_c}(b)).
As shown in \cref{fig:reheat}(b), $A_4$ varies immensely (by three orders of magnitude),
mostly due to the variation of the temperature $T_a$,
with only a mild dependence on the time $t_a$.
Moreover, we show in \cref{fig:reheat}(c) that these curves can be collapsed,
in the range of parameters probed, assuming the functional form
${A_4(T_a,t_a)=f (t_a^\gamma \, \exp(-E_a / T_a))}$
and ${\gamma = 0.2}$.
The function $f$ is linear
at small argument, supporting an Arrhenius behaviour at low temperature~$T_a$
(see \cref{sec:qlm}).
Remarkably, this collapse indicates that for a given gap,
the distribution of excitation energies is characterised by a single energy scale $E_a$
(presumably a lower cutoff, see below).

Interestingly, we find in \cref{fig:reheat}(d) that $E_a$ very strongly increases with
gap magnitude $\omega_c$ (see below for a proposed explanation).
The dynamical exponent $\gamma(\omega_c)$  is also shown in inset,
and remains smaller than $0.2$ in the entire range of initial gaps that we probe.

\section{Modes beyond the gap act as an excitation reservoir}

We saw that, if we start from a glass with an initially gapped density of states, thermal fluctuations will always populate this gap.
To rationalise these findings, we consider the path of
minimal energy connecting two states associated to one excitation,
and denote by $s$ the curvilinear coordinate along it.
The Taylor expansion of the energy along this path
from the state 1, by definition the one of minimal energy, reads:
\begin{equation}
    \label{eq:Taylor}
    E(s)
    = \frac{1}{2!} \lambda_1 s^2
    + \frac{1}{3!} \kappa_1 s^3
    + \frac{1}{4!} \chi_1 s^4+{\cal O}(s^5)
\end{equation}
which is a double-well, with a curvature  ${\lambda_1 \approx \omega_1^2}$ around the minima in state 1. Physically, $\chi_1>0$ (otherwise the potential has no lower energy limit).

In that formalism, starting from a gapped glass corresponds to having a distribution ${P(\lambda_1,\kappa_1,\chi_1)}$ strictly zero at $\lambda_1 < \omega_c^2$ and smooth above $\omega_c^2$.
At finite temperature the gap is populated by thermal activation towards a state 2 with a smaller frequency ${\omega_2\approx \sqrt{\lambda_2}}$, which corresponds to a transition in an asymmetric double-well (as illustrated in Fig.~\ref{fig:intro}).
From \cref{eq:Taylor} it is straightforward to obtain the expansion from state 2,
and the transformation ${(\lambda_2, \kappa_2, \chi_2) = g(\lambda_1, \kappa_1, \chi_1)}$.
The joint distribution follows
${P(\lambda_2,\kappa_2,\chi_2)=|g'(\lambda_2,\kappa_2,\chi_2)| P(\lambda_1,\kappa_1,\chi_1)}$
where the absolute value of the determinant of the Jacobian ${\vert g'(\lambda_2,\kappa_2,\chi_2) \vert \sim\lambda_2}$
for small $\lambda_2$ (see \cref{sec:Jacobian}).
Owing to the smoothness of $P(\lambda_1,\kappa_1,\chi_1)$ for ${\lambda_1\gtrsim \omega_c^2}$, for small $\lambda_2$ one has
$P(\lambda_2,\kappa_2,\chi_2)\sim \lambda_2$
or equivalently
$P(\omega_2,\kappa_2,\chi_2)\sim \lambda_2 d\lambda_2/d\omega\sim\omega_2^3$.
After integrating on $\kappa_2$ and $\chi_2$ one gets ${D_L(\omega_2)\sim\omega_2^3}$.
See \cite{Gurarie03} for a more general argument along the same line.
Thus one expects to observe a pseudo-gap
following thermally activated excitations.
One effect will deplete the spectrum even further:
in the case of a cubic pseudo-gap, the low-frequency spectrum is dominated
by states 2 very close to a saddle node bifurcation (at the spinodal).
However, once interactions among excitations are taken into account\footnote{%
    Such interactions are relevant even at high temperature
    near the glass transition \cite{Lemaitre14}.
    Note that when quenching the system
    to zero temperature after a reheating, interactions with relaxing vibrational modes
    may also destabilise excitations which are close to their spinodal.
},
configurations with such a large density of states near saddle node bifurcation can be shown
to be unstable and display avalanche-type events,
\textit{i.e.}~where the relaxation of one excitation can
destabilise others in turn \cite{Wencheng19}.
This effect will increase the pseudo-gap exponent to values larger than three\footnote{%
    Interactions between excitations cause a pseudo-gap in the density $P(x)\sim x^\theta$ of excitations
    within a force $x$ to fail \cite{Lin14a,Lin16}.
    Near a saddle-node bifurcation, one has $\omega \sim x^{1/4}$ leading to
    $D_L(\omega)\sim \omega^\alpha$ where $\alpha=3+4 \theta$.
    Empirically $\theta\approx 0.3$ after a slow quench from high temperature
    (but it is larger for a fast quench) \cite{Wencheng19},
    which would lead to an exponent ${\alpha\approx 4.2}$ consistent with our measurement.
}.

As far as the kinetics is concerned, the time scale $t_a$ on which an excitation equilibrates
depends on the energy barrier $\Delta E$ to go from state 1 to 2.
It will occur (neglecting prefactors) when ${t_a \gg t_a^*\sim \exp(\Delta E/T)}$,
\textit{i.e.}~its first-passage time.
For much larger time scales, the probability of being in the
excited states follows a Boltzmann factor
$\exp(- E_{12} /T)$ at small $T$, where $E_{12}$ is the energy difference between the two states.
If all states were equilibrated, $A_4$ would not depend on $t_a$ (\textit{i.e.}~$\gamma=0$).
By contrast, if no states were equilibrated $A_4$ would grow linearly in time.
In that respect, our observation of the intermediate case ${\gamma \approx 0.2}$
is consistent with the notion that
there is a broad distribution of barriers,
so that on the time scale $t_a$ a fraction of excitations are equilibrated,
yet some barriers are still being jumped over for the first time.

For a given gap magnitude $\omega_c$, we expect to find a lower cutoff on the
distribution of barriers $\Delta E$ (with the typical energy difference $E_{12}$ of the associated excitations being of the same order of magnitude).
Consider for instance a symmetric double-well
in the energy landscape and expand its energy around the maximum:
${E(s)=- \frac{1}{2!} \lambda s^2 +\frac{1}{4!} \chi s^4}$.
It is straightforward to show that in each minimum the frequency of the soft mode scales as ${\sqrt \lambda}$,
allowing us to identify for the softest excitations ${\lambda\sim \omega_c^2}$.
Likewise in this example the barrier for the double-well follows
${\Delta E\sim \lambda^2\sim \omega_c^4}$.
This scaling holds for asymmetric double-wells as well (see Appendix E).
Interestingly, our measured activation energy $E_a$ is compatible with this power-law relation,
except for the smallest gap (hence less stable glass) that we study\footnote{%
    We observed that for small gaps the assumption that all the excitations
    are in their energy minimum breaks down,
    but on the contrary for our largest gap it holds true for about $90\%$ of excitations causing
    quasi-localised modes.
    This state of affairs is expected since for very small gaps particles can hardly breathe,
    and the corresponding gapped states obtained by our protocol are not extremely stable.
}.

Overall, this analysis supports the scenario that modes beyond the gap act
as a reservoir of excitations,
with a broad distribution of barriers presenting a typical cutoff
$E_a\sim \omega_c^4$ at low energies.

\section{Effects of a thermally filled-up gap on physical properties}

\subsection{The softening of loading curves is proportional to \texorpdfstring{$A_4$}{A4}}

We now discuss the practical implications of the preparation dependent amplitude ${A_4(T_a,t_a)}$ on the mechanical properties of the ultra-stable glass. The relationship between shear transformation and quasi-localized modes was studied in \cite{Wencheng19} for rapidly quenched glasses. More generally, more stably prepared systems exhibit a steeper loading curve and have a lower density of quasi-localised modes \cite{Ozawa18, Wencheng19}. Here, we show a quantitative relationship between the amplitude of quasi-localised modes ${A_4}$ and the effective shear modulus during loading ${\mu \equiv \langle\Sigma \rangle/\epsilon}$, where $\epsilon$ is the imposed shear strain and $\langle \Sigma \rangle$ is the ensemble average of the corresponding shear stress increase \footnote{Shear stress is measured relative to the initial value at $\epsilon = 0$.}. More plasticity leads to a smaller $\mu$.

\begin{figure}[ht]
	\centering
	\includegraphics[width=\linewidth]{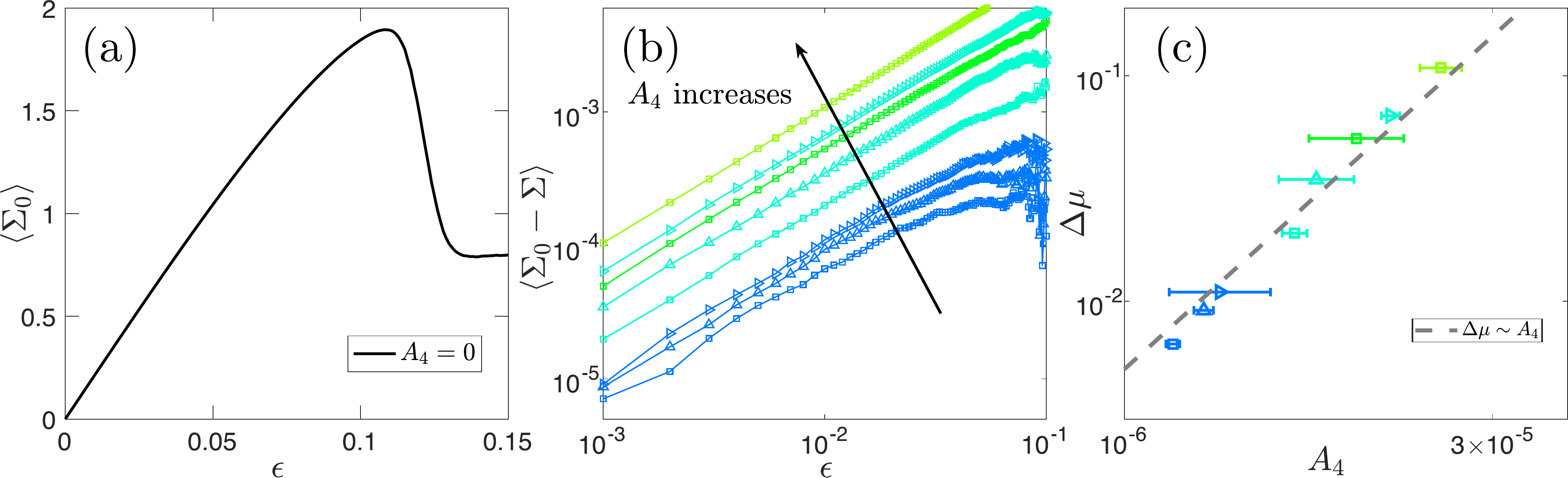}
	\caption{
	    (a)~The ensemble averaged shear stress $\langle \Sigma_0 \rangle$ as a function of strain $\epsilon$ for our largest gap ($\omega_c = 1.64$).
		(b)~The difference $\langle\Sigma_0-\Sigma\rangle$ as a function of strain $\epsilon$ for different temperature cycles applied to the gapped glass from (a). As observed, the stress decreases as $A_4$ increases (from bottom to top,  $T_a=\{0.15,0.15,0.15,0.2,0.2,0.2,0.23,0.25\}$ corresponding to $t_a=\{5\times10^2,2\times10^3,10^4,5\times10^2,2\times10^3,10^4,5\times10^2,5\times10^2\}$, respectively).
		(c)~The decrease in the effective shear modulus $\Delta\mu$ computed at ${\epsilon=10^{-2}}$ is proportional to $A_4$: the dashed line corresponds to $\Delta \mu \sim A_4$.}
	\label{fig:A4_stress}
\end{figure}

We measure the stress-strain response in ultra-stable glasses using a quasi-static loading protocol (see \cref{fig:A4_stress}(a)). We find that the gapped glasses have the highest effective shear modulus ${\mu_0= \langle\Sigma_0\rangle/\epsilon}$ and it decreases as the gap is filled and $A_4$ increases (\cref{fig:A4_stress}(b)). The reduction of the effective shear modulus ${\Delta \mu(A_4) \equiv \langle \Sigma_0 - \Sigma(A_4)\rangle/\epsilon}$ is proportional to $A_4$ (see \cref{fig:A4_stress}(c)) in the range of strains $\epsilon < 0.01$ where it is strain independent.

\subsection{TLS disappear for large gaps}
We argue that TLS cannot be observed if a glass presents a large gap. Indeed if the tunneling amplitude is too small,  on experimental timescales a single state is visited and TLS properties are not apparent \cite{Parshin07}. It is precisely what happens when the gap is large, as barriers are then both larger and wider. To estimate this effect  we follow the treatment of soft potential models \cite{Parshin93} that solves the Schr\"{o}dinger equation in potentials described by \cref{eq:Taylor}. For a symmetric double-well, the tunneling time follows $\tau=\hbar \pi/\Delta_0$, where  $\Delta_0$ is the splitting energy stemming from quantum tunneling (and $\hbar$ is the reduced Planck constant).  $\Delta_0$ is expressed as $\Delta_0=W\exp(- (\omega_1 / \bar{\omega})^3)$  where $W= \hbar (\hbar\chi_1/(96m^2))^{1/3}$ and $\bar\omega=(\hbar\chi_1/(2m^2))^{1/3}$  (cf.\ \cite{Parshin93}), with $m$ the particle mass.
Thus if ${\omega_1>\omega_c^*\equiv(\ln \frac{\tau W}{\hbar\pi})^{1/3}\bar\omega }$, TLS are not apparent.
In our simulations we find that the median of $\chi_1 \approx 4.6 m \omega_{D}^{2}/a^{2}$ (for our largest gap) where $a$ is the inter-particle distance and $\omega_D$ is the Debye frequency (see \cref{sec:estimation} for details). Taking estimates in amorphous silicon where $\omega_D\approx k_B 530 \text{K} /\hbar$ \cite{Mertig84}, we get $W\approx0.03\hbar\omega_D$ and $\omega_0\approx0.1\omega_D$. Considering the experimentally accessible time scale to be of order $\tau\approx 100 \text{s}$, at last we estimate $\omega_c^*\approx0.3 \omega_D$.
It is of the order of magnitude of our largest gap $\omega_c\approx0.1\omega_D$.

Suppressing TLS altogether would thus be  accomplished by preparing sufficiently stable glasses so as to get $\omega_c > \omega_c^*$.
These considerations stay valid even when thermal activation populates the gap and $A_4$ becomes finite,
because quasi-localised modes with low-frequency correspond then to quite asymmetric wells whose barrier height and width (and therefore tunneling amplitude) is still comparable to the estimate above. The presence of a large underlying, thermally populated, gap of quasi-localised modes in ultrastable glasses thus offers an explanation  for their lack of TLS \cite{Queen13,Perez14}.

\section{Rearrangements involve fewer particles and become string-like at large gaps}

We introduce a novel algorithm to decompose the displacement field of a rearrangement
into several elementary excitations, which is needed to study how their
geometry depends on the gap magnitude.
Given a displacement field (induced, in our case, by the thermal cycle),
we first consider the particle with maximal displacement,
and draw a sphere of radius $\tilde{R}$ around it.
Beyond this sphere, all the particle displacements are set to zero
(\textit{i.e.}~these particles are set back to their initial position in the gapped state),
whereas within the sphere the displacements are preserved.
Next, we perform with that initial condition a steepest descent
of the interaction energy.
We find that if $\tilde{R}$ is small, all displacements go back to zero,
whereas if $\tilde{R}$ is large, they do not.
We consider the smallest $\tilde{R}$ of the latter case,
and the displacement field obtained at the end of the corresponding gradient descent
defines our first elementary excitation.
Next, we subtract this obtained displacement field from the full one,
and repeat the entire procedure recursively until no more excitations are found
(see \cref{sec:geometry} for details and visual examples).
\begin{figure}[ht]
\centering
    \includegraphics[width=\linewidth]{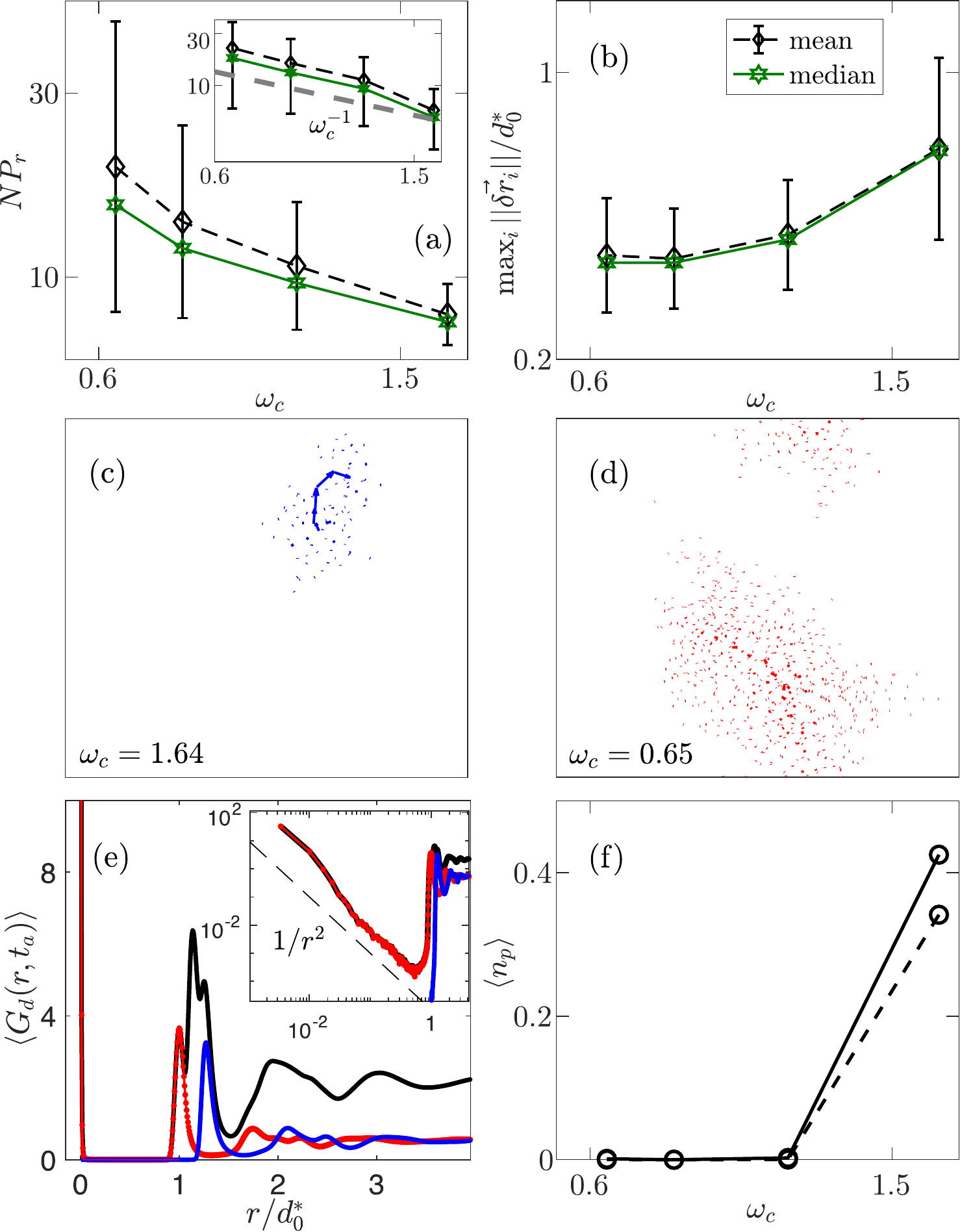}
    \caption{
        (a)~Mean (dashed black) and median (green) of the number of particles $N P_r$ involved
        in an individual excitation \emph{vs} $\omega_c$ at $t_a=500$
        (at the lowest temperature we probe for each gap).
        The error bars stand for the standard deviation of $N P_r$ which indicates that
        the distribution of $NP_r$ is broader for smaller $\omega_c$.
        (b)~Mean (dashed black) and median (green) of the displacement norm
        of the most mobile particle in an excitation, as a function of $\omega_c$.
        $d_0^*$ is the most frequent diameter
        of the smaller particles
        ($d_0^*=\{0.738,0.918,0.963,0.988\}$ for the different $\omega_c$).
        (c)-(d)~Thermally induced rearrangement, respectively for our largest ($\omega_c=1.64$)
        and smallest ($\omega_c=0.65$) gaps, projected on the $xy$ plane.
        (e)~Ensemble and
        radially averaged (on all observed excitations) Van Hove correlation function
        ${G_d (r, t_a)}$ computed for all (in black and having largest $G_d$ at large $r$), only small (red dotted)
        or only large (blue) particles for ${\omega_c = 1.64}$.
        The peak around ${r = 0}$ corresponds to permutations of particles.
        The black and red curves overlap, indicating that permutations only
        occur on small particles.
        (f)~Average number of permutations $\langle n_p \rangle$
         \emph{vs} $\omega_c$,
        for two different cutoff distances ${r_c / d_0^* = \{0.025, 0.05\}}$
        (dashed and solid, respectively).
    }
    \label{fig:geometry}
\end{figure}

Given an individual excitation of displacement field $\{\delta \vec{r}_i\}$,
we compute, from its associated participation ratio, an estimate of
the number of particles involved in this excitation
${N P_r \equiv [\sum _i || \delta \vec{r}_i ||^2]^2/\sum _i || \delta \vec{r}_i ||^4}$.
For each gap magnitude $\omega_c$,
we find about 5000 such excitations and report the mean and the median
of this observable in \cref{fig:geometry}(a).
We find that the typical number of particles involved in
one excitation decreases as $\omega_c$ increases.
We propose the following rationalisation.
The length scale of quasi-localised modes was found to be proportional to
the characteristic length scale entering the response to a local dipole \cite{Shimada18},
as proposed based on a variational arguments in \cite{Yan16}.
The length scale $\ell_c$ entering the dipolar response was observed to decrease
as the system moves away from a marginally stable phase and
enters a gapped solid phase as
${\ell_c \sim 1 / \sqrt{\omega_c}}$ \cite{Lerner14}.
The volume of the corresponding mode was shown to go as $\ell_c^2$
(independent of the number of dimensions)
for elastic networks of springs at rest \cite{Lerner14,Yan16}.
Taken together, these results  correspond  to a
number of particles involved in an excitation  that decreases with increasing gap
as $N P_r \sim 1/\omega_c$.
As shown in the inset of \cref{fig:geometry}(a), this is in reasonable
agreement with our observations.

A complementary observable for the geometry is the maximum displacement
${\max_i \{|| \delta \vec{r}_i ||\}}$ for a given excitation,
whose mean and median values for all excitations at a given $\omega_c$
are shown in \cref{fig:geometry}(b).
Interestingly, this maximum displacement increases with the gap,
and becomes close to the small particle diameter
equal to $d_0^*$.

A direct visualisation of the excitation fields reveals a
(presumably related) interesting phenomenon:
for our largest gap, the displacements are string-like with
several particles exchanging positions,
as shown in \cref{fig:geometry}(c), whereas for smaller gaps they are much
more compact and no permutations occur.
To quantify this effect, we follow the glass transition literature
\cite{Donati98} and measure the distinct part of the Van Hove correlation:
\begin{equation}
    G_d (\vec{r}, t) \equiv \frac{1}{N} \left\langle
        \sum\limits_{i = 1}^N
        \sum\limits_{j (\neq i)}^N
        \delta \left(
            \vec{r} - \vec{r}_j (t) + \vec{r}_i (0)
        \right) \right\rangle ,
\end{equation}
where the average is made on all the observed elementary excitations at some given $\omega_c$.
It is plotted for our stablest system in \cref{fig:geometry}(e) after radially averaging.
The key observation is the presence of a very sharp peak around $r = 0$,
which can only arise from particles replacing each other.
Interestingly, if we condition our definition of the Van Hove correlation
to large or small particles only, we find that the peak only persists for small particles
(in red in \cref{fig:geometry}(e)).
Strings thus correspond to smaller particles navigating in an environment of larger ones.

Next we integrate the peak around $r = 0$ to quantify the number of permuting
particles averaged on all elementary excitations:
\begin{equation}
    \langle n_p \rangle =
        \left\langle N \int_0^{r_c} G_d(r, t_a) \, 4 \pi r^2 dr \right\rangle ,
\end{equation}
where $r_c$ is a cutoff that is tuned.
We observe that permutations are essentially absent except for the largest considered gap,
see \cref{fig:geometry}(f).

\section{Discussion}

In summary, we have argued that in gapped glasses,
modes beyond the gap act as an excitation reservoir for thermal activation.
This effect always destroys the gap and leads to a density of quasi-localised modes
${D_L(\omega)\approx A_4(T) \, \omega^4}$.
At low temperatures, we found that ${A_4\sim \exp(-E_a(\omega_c)/T)}$ where the typical energy scale
$E_a(\omega_c)$ is a rapidly increasing function of $\omega_c$.
A simple scaling prediction gives $E_a \sim \omega_c^4$,
in good agreement with our observations,
except for the smallest gap value that we explore.
Finally, we observed that as the gap increases, excitations involve fewer and
fewer particles and become more and more string-like.
The growing length scale of the excitations as $\omega_c \rightarrow 0$
is consistent with the previously identified growing length
characterising the elastic response of an amorphous solid near
a macroscopic elastic instability \cite{Lerner14}.

Although our observations were made in ultra-stable states obtained by a specific protocol\footnote{
    In general to create gapped glasses one needs to use a gradient descent method using swap or equivalently breathing particles up to zero temperature. If instead one uses MD with the normal dynamics to quench from a finite temperature, the gap of magnitude $\omega_c$ will be filled by excitations},
our arguments on this reservoir effect are much more general.
Assuming that this effect is at play in supercooled liquids ties together
several unexplained observations, as we now discuss.

{\it Reinterpreting mean-field descriptions of glasses:}
Goldstein \cite{Goldstein69} proposed early on that the glass transition takes
place near some temperature $T_c$ below which most normal modes become stable.
Such an enhanced stability is consistent with the overall elastic stiffening upon cooling
apparent in the bulk \cite{Hecksher15} or local \cite{Lerner18}
elastic moduli in fragile supercooled liquids.
Theoretically, this view is consistent with mean-field models of
the glass transition in infinite dimensions --
that are closely related to Mode-Coupling Theory --
\cite{Lubchenko07,Biroli12},
in which the spectrum of the Hessian becomes stable and opens
a gap with ${\omega^2_c \sim (T_c - T)}$ \cite{Parisi03}.
Our work suggests a natural way to extend this picture to finite dimensions
as sketched in \cref{fig:intro}:
the gap is decorated by excitations stemming from the
reservoir of modes with $\omega\geq \omega_c$.
In this approach
\textit{(i)}~the excitation density strongly decreases
with temperature: away from $T_c$ in the deeply supercooled regime,
it should be proportional to $\exp(-E_a(\omega_c(T))/T)$ and
\textit{(ii)}~as $T$ decreases, $\omega_c$ increases and excitations are less and less extended.
Point~\textit{(i)} offers an explanation for the very rapid decay
upon cooling of $A_4(T)$ \cite{Wang19,Rainone20},
TLS density \cite{Khomenko19} and shear transformations \cite{Ozawa18,Jin18}
observed in ultra-stable supercooled liquids.
Point~\textit{(ii)} is consistent with the result that TLS \cite{Khomenko19}
and quasi-localised modes \cite{Wang19,Rainone20} present
a lower participation ratio upon cooling (such changes of
geometry may lead to additional effects on their density\footnote{%
    See \cite{Rainone20} for a discussion on quasi-localised modes.
    Concerning TLS, a smaller participation ratio suggests a higher tunnelling amplitude,
    which may in turn affect the TLS density.}
).

{\it Glass transition:}
The mean-field proposal that supercooled liquids present an effective gap growing upon cooling,
leading to a rarefaction of thermally accessible excitations,
is consistent with the observation that rearrangements become string-like
with more and more particles exchanging positions upon cooling \cite{Donati98,Yu17}
-- since we find that excitations at large gap are precisely like that.
In our view, why elementary excitations display such a geometry
at large gap is yet to be explained\footnote{%
    It was argued within RFOT that strings would exist at
intermediate temperatures,
    \cite{Stevenson06,Stevenson10}.
    Yet, this analysis is based on the ``library of states'' picture that describes only
    rough barriers consisting of many intermediary ones (that must exist between each state of the library), leading to impact rearrangements.
    This description thus cannot apply to the elementary excitations
    displaying a single barrier studied here. This picture is explicitly different than our view.
    In our view, a description of strings should also explain why particles exchange positions.
}.

At our lowest temperatures, only one or a few strings get activated,
which can only lead to a very partial
relaxation of the system.
Isolated strings thus belong to the class of
$\beta$-relaxation in supercooled liquids,
as shown in \cite{Yu17}, using normal dynamics, for model metallic glasses
(molecular and covalent liquids may certainly present other $\beta$-relaxation mechanisms
governed by the local chemistry, such as dangling bonds).
Yet  strings may also contribute the $\alpha$-relaxation
of liquids if they are present in sufficient density, at least for the continuously polydisperse
ones receiving much attention currently.
Indeed they allow for the exchange of particles with distinct radii.
Such swap moves are now known to relax the system with great efficiency,
so the dynamics should not be slower than the time scale to naturally operate these swaps
\cite{Wyart17,Ikeda17}. The rapid increase of their characteristic low-energy cutoff $E_a$
with growing gap would then contribute to the fragility of liquids.

Note that such  views in which activation deep in the supercooled liquid phase is
controlled by $T_c$, contrasts with the usual interpretation of mean-field results in which
activation is controlled by an entropy crisis
occurring at a lower temperature $T_K$ (the Kauzmann temperature) \cite{Lubchenko07,Biroli12}.
The latter is in our opinion ruled out in poly-disperse systems by the recent observation that
changes of kinetic rules (such as allowing for swap moves \cite{Ninarello17})
immensely affect the location of the glass transition,
while leaving intact thermodynamic properties \cite{Wyart17}.
Changes of kinetic rules, however, affect the location of the mode coupling temperature $T_c$ \cite{Ikeda17,Brito18},
thus pictures of activation based on that temperature are consistent with the observations of swap algorithms.

{\it Effects of rare fluctuations:} We have shown that a gap in the density quasi-localised modes cannot exist in finite dimension at finite temperature, due to  the thermal activation of their associated excitations. We expect that at least another effect will enter in finite dimension to fill up the gap. In electronic systems that present impurities, the density of states does not vanish in the range of energies where the pure system would, due to rare regions where many impurities are present. This effect leads to the so-called ``Lifshitz tail'' in the electronic density \cite{Lifshitz65}. In glasses,  we expect that aspects of the structure controlling stability, such as coordination and pressure \cite{Alexander98,Wyart05a}, will also fluctuate and lead to rare weaker regions in the materials (we do not see this effect in our breathing particles, whose preparation may lead to an unusually homogeneous material). Such fluctuations will need to be larger and larger as the gap grows to contribute to low-frequency quasi-localised modes, and therefore less likely, leading to a rapidly decaying density of quasi-localised modes with growing gap. These atypical rare regions may have little effect for plasticity or structural relaxation near the glass transition, but may be important in affecting the density of TLS. It would thus be interesting in the future to study glasses of controlled inhomogeneity to separate rare fluctuations in the structural disorder from the `excitations reservoir' effect introduced here.

\section*{Acknowledgements}

We would like to thank L.~Berthier, G.~Biroli, C.~Cammarota, D.~Khomenko, C.~Liu,
V.~Lubchenko, M.~M{\"u}ller, M.~Ozawa, C.~Scalliet, P.~Wolynes
and F.~Zamponi for fruitful discussions,
as well as G.~Kapteijns and E.~Lerner for discussions and numerical support at the beginning.
T.G.~acknowledges support from The Netherlands Organisation for Scientific Research (NWO)
by a NWO Rubicon Grant 680-50-1520 and from the Swiss National Science Foundation (SNSF)
by the SNSF Ambizione Grant PZ00P2{\_}185843.
E.A.~acknowledges support from the SNSF by the SNSF Ambizione Grant PZ00P2{\_}173962,
and M.W.~by the Simons Foundation Grant (\#454953 Matthieu Wyart) and
from the SNSF under Grant No.\ 200021-165509.

\bibliographystyle{unsrtnat}
\bibliography{library}

\clearpage
\onecolumn
\justify
\appendix

\section{Parameters \& Key quantities}
\label{sec:nomenclature}

\begin{center}
	\renewcommand{\arraystretch}{1.2}
	\newcolumntype{C}[1]{>{\centering\let\newline\\\arraybackslash\hspace{0pt}}m{#1}}
	\begin{tabular}{|C{8mm}||C{33.5mm}|C{33.5mm}|C{33.5mm}|C{33.5mm}|}
		\hline
		$K$ & $10^2$ & $10^3$ & $3\times10^3$ & $10^4$ \\
		\hline
		$\omega_c$ & $1.65$ & $1.19$ & $0.85$ & $0.65$ \\
		\hline
		$n$ & $4000$ & $4000$ & $2000$ & $3000$ \\
		\hline
		$N$ & $8000$ & $8000$ & $8000$ & $8000$ \\
		\hline
		$\phi$ & $0.62$ & $0.74$ & $0.77$ & $0.79$ \\
		\hline
		$\langle V\rangle$ & $3722.0\pm0.3$ & $7303.9\pm0.7$ & $8607.2\pm0.9$ & $9356\pm1$ \\
		\hline
		$\rho$ & $2.15$ & $1.10$ & $0.93$ & $0.86$ \\
		\hline
		$T_p$ & $0.2$ & $0.39$ & $0.44$ & $0.49$ \\
		\hline
		$t_p$ & $1e4$ & $1e4$ & $1e4$ & $1e4$ \\
		\hline
		$T_c$ & $0.304$ & $0.484$ & $0.525$ & $0.572$ \\
		\hline
		$T_a$ &
		$\{0.15, 0.20, 0.23, $ &
		$\{0.07,0.10, 0.20, 0.30, $ &
		$\{0.03, 0.05, 0.1\}$ &
		$\{0.03, 0.05, 0.10, 0.20, $ \\
		&
		$\hphantom{\{} 0.25, 0.27, 0.30, 0.40\}$ &
		$\hphantom{\{} 0.39, 0.50, 0.60, 0.80\}$ &
		&
		$\hphantom{\{} 0.30, 0.49, 0.60, 0.80\}$ \\
		\hline
		$t_a$ & $\{500, 2000, 10000\}$ & $\{100, 500, 2000\}$ & $\{500\}$ & $\{100, 500, 2000\}$ \\
		\hline
		$E_a$ & $0.66\pm0.08$ & $0.15\pm 0.03$ & $0.042\pm0.016$ & $0.03\pm0.007$ \\
		\hline
		$\gamma$ & $0.2$ & $0.05$ & - & $0.003$ \\
		\hline
		$\langle n_e \rangle$ & $0.65$ & $0.89$ & $1.2$ & $3.2$ \\
		\hline
	\end{tabular}
\end{center}

\noindent Notice that $\phi\equiv \sum_{i=1}^N \tfrac{4}{3} \pi R_i^3 / V$ 	is the packing fraction;  and that $\langle V \rangle$
is the ensemble average volume (and the uncertainty its standard deviation). The average number of excitations per realisation $\langle n_e \rangle$ has been obtained with $t_a = 500$ at the lowest $T_a$ we probed for each $K$. $T_c$ is obtained from the relation: relaxation time  $\sim (T-T_c)^{-\nu}$ \cite{Cavagna09}, where $\nu$ is also a fit parameter. Lengths ($\vec{r}$, $R_i$ and $L$) are shown in the unit of $d_0$ (or the most frequent diameter of small particles $d_0^*$, if so indicated):
the diameter of an initially small particle. Energies ($E_a$) are expressed in the unit of $\varepsilon$ that is the prefactor of the pair interaction potential (see below). Temperature ($T_p$, $T_a$, and $T_c$) is in the unit of $\varepsilon / k_B$. Time ($t_p$, $t_a$, and $\omega^{-1}$) is shown in the unit of
$t_0$ where $t_0 \equiv \sqrt{m d_0^2 / \varepsilon}$.

\subsection*{Detecting a rearrangement}

To detect if reheating with a temperature $T_a$ for a duration of $t_a$ has led to a rearrangement
we consider the ratio of the norm and the participation ratio of the displacement field:
\begin{equation}
\mathcal{I} \equiv \frac{|| \delta \vec{r}_i ||}{P_r(\delta \vec{r}_i)}
\end{equation}
where $|| \delta \vec{r}_i ||$ is the Euclidean norm of the particle displacement field, $\delta \vec{r}_i$, between the
quenched states before and after reheating, and its participation ratio
\begin{equation}
P_r (\delta \vec{r}_i) \equiv
\frac{(\sum_i || \delta \vec{r}_i ||^2)^2}{N\sum_i || \delta \vec{r}_i ||^4}
\end{equation}
If a rearrangement results from reheating, the norm of the displacement field is finite
(typically $10^{-3}$ -- $10^{-2}$) and participation ratio in the order of $10^{-3}$ -- $10^{-2}$.
In contrast, if there was no rearrangement, the norm of the displacement field
is of the order of the numerical precision ($10^{-6}$) and the participation ratio is of order one.
We distinguish the two cases using a threshold.
We define that a rearrangement has taken place if $\mathcal{I} > 10^{-3}$.
Note that the distributions of $\mathcal{I}$ corresponding to the two cases are clearly separated.

\section{Molecular dynamics}
\label{sec:md}

\subsection*{Sample preparation: `breathing' dynamics}
\label{sec:md:swap}

We study a three-dimensional periodic particle system of ${N = 8000}$ particles,
that is characterised by the grand potential
\begin{equation}
\mathcal{U}
= \sum_{i<j} \varphi \left( r_{ij},\, R_i,\, R_j \right)
+ \sum_i \mu \left( R_i,\, R_i^{(0)} \right)
\end{equation}
where $\varphi$ is a purely repulsive inverse power-law potential, defined
\begin{equation}
\label{eq:intU}
\varphi \left( r_{ij},\, R_i,\, R_j \right) =
\begin{cases}
\displaystyle
\varepsilon \left[
\left( \frac{R_{ij}}{r_{ij}} \right)^{10} +
\sum_{p=0}^3 c_{2p} \left( \frac{r_{ij}}{R_{ij}} \right)^{2p}
\right],
\quad &
\displaystyle
\frac{r_{ij}}{R_{ij}} \leq r_c
\\
0,
\quad &
\displaystyle
\frac{r_{ij}}{R_{ij}} > r_c
\end{cases}
\end{equation}
with $r_c$ is the cutoff distance,
$R_{ij} \equiv R_i + R_j$ (two times the average particle radius), and
$r_{ij} \equiv || \vec{r}_{ij} || \equiv || \vec{r}_i - \vec{r}_j ||$
(the Euclidean norm of the distance vector
separating particles $i$ and $j$).
$c_{2p}$ is a constant that makes $\varphi$ continuous up to the third derivative at $r_c$.
Furthermore,
\begin{equation}
\mu \left( R_i,\, R_i^{(0)} \right) =
\frac{K}{2} \left( 1 - \frac{R_i^{(0)}}{R_i} \right)^2 \left( R_i^{(0)} \right)^2
\end{equation}
is a chemical potential that allows a particle to change its size from its initial value $R_i^{(0)}$
at an energetic cost that scales with a modulus $K$.
For $K = \infty$ it is impossible for a particle to change its radius,
while it becomes easier as $K \rightarrow 0$.
The initial particle radii are bi-disperse, in a 50:50 mixture.
In particular, one, randomly selected, half of the particles has $R_i^{(0)} = 0.5 d_0$ and
the other half has $R_i^{(0)} = 0.7 d_0$
(where $d_0$ sets the unit of length of our system).

Sample preparation proceeds by instantaneously heating the initial random configuration
to a temperature $T_p$ and keeping it at this temperature for a certain time $t_p$
under the constraint of a fixed pressure $p=20.0$ (in units of $\varepsilon / d_0^3$).
We then instantaneously quench the system to zero temperature by minimising the grand potential.
See algorithmic details below.

\subsection*{Activation by temperature: normal dynamics}
\label{sec:md:normal}

We proceed by fixing the particle size, which corresponds to a potential energy
\begin{equation}
U = \sum_{i<j} \varphi \left( r_{ij},\, R_i,\, R_j \right)
\end{equation}
(see \cref{eq:intU} for the definition of $\varphi$).
We then gently heat the system configuration
to a certain ``activation temperature'' $T_a$
(at a heating rate $T_a / (10 \, t_0)$),
and keep the sample at $T_a$ for a total duration $t_a$.
Thereafter we instantaneously quench the sample to zero temperature.
Algorithmic details are listed below.

\subsection*{Molecular dynamics algorithm}
\label{sec:md:algorithm}

We run molecular dynamics, whereby the particle dynamics are given
by Newton's equation of motion with the gradient of the potential
energy on a particle as driving force.
Time is discretised in steps of $\Delta t$ using the standard velocity Verlet algorithm.
The temperature and pressure are controlled using
a Berendsen thermostat \cite{Berendsen84},
where the temperature is defined as the total kinetic energy
$\sum_i m || \dot{\vec{r}}_i ||^2 / 2$
(where $\dot{\bullet}$ refers to the time derivative).
Note that during preparation the kinetic energy
is $\sum_i (m || \dot{\vec{r}}_i ||^2 + \dot{R}_i^2) / 2$.
We use the `FIRE' algorithm \cite{Bitzek06} to quench the systems.

For completeness we report that $r_c = 1.48 \, d_0$, $\varepsilon =1$, $m =1$, $d_0 = 1$,
and $\Delta t = 0.005$.
Furthermore
$c_0 = -1.1106337662511798$,
$c_2 = 1.2676152372297065$,
$c_4 = -0.4960406072849212$,
$c_6 = 0.0660511826415732$;
see Supplemental Material of \cite{Lerner17}.

\section{Sample preparation}
\label{sec:sample}

\begin{figure}[ht]
	\centering
	\includegraphics[width=0.6\linewidth]{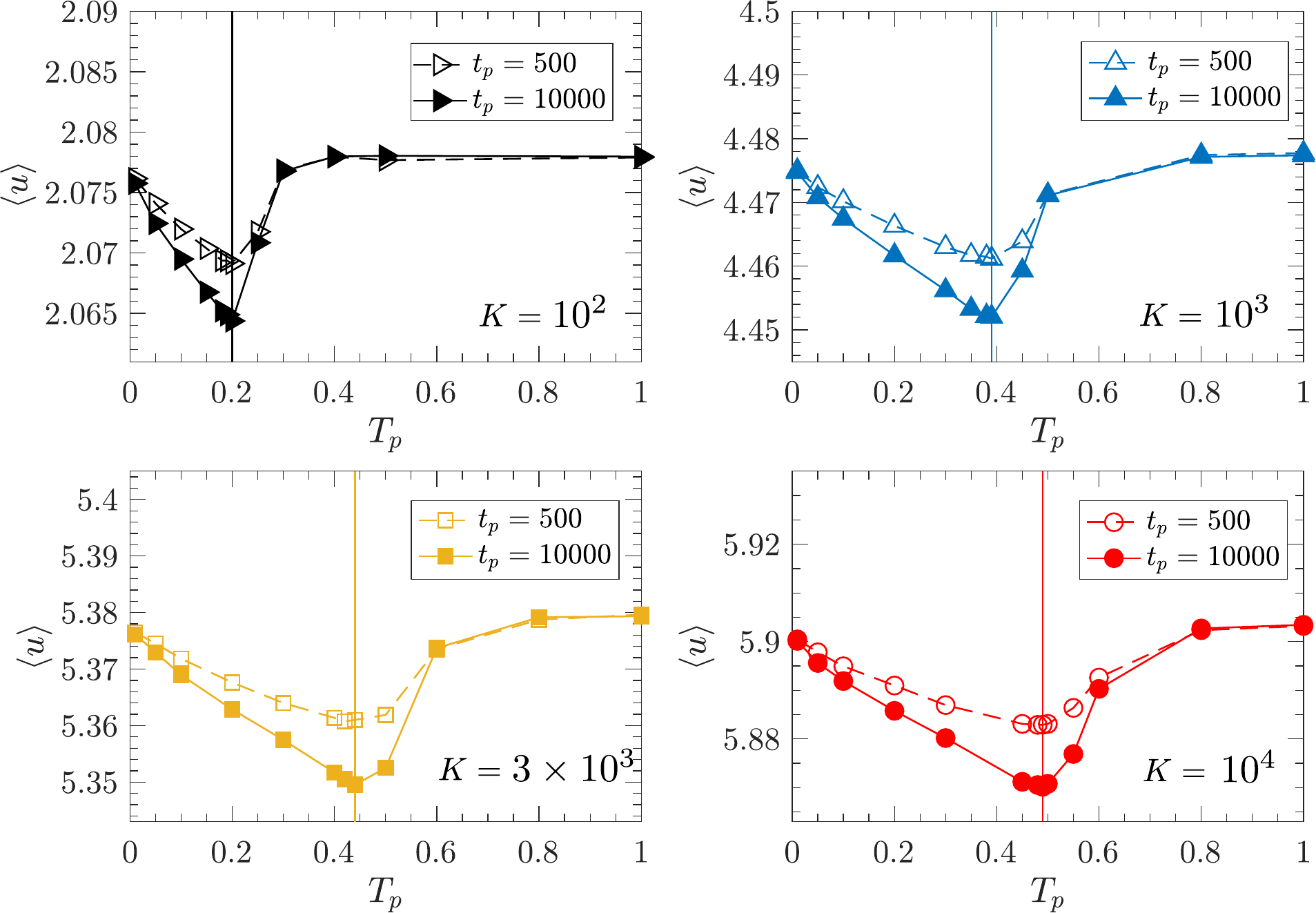}
	\caption{
		Mean interaction potential energy $\langle u\rangle$ after
		sample preparation with `breathing' dynamics for varying
		parent temperature $T_p$ and two different waiting times $t_p$.
		The different panels correspond to different $K$ as indicated.
		The selected temperature $T_p$ for which the potential energy is
		lowest for the largest practically reachable $t_p = 10^4$
		is indicated using vertical lines
		(see \cref{sec:nomenclature} for numeric values).}
	\label{fig:Tp}
\end{figure}

\begin{figure}[ht]
	\centering
	\includegraphics[width=0.33\linewidth]{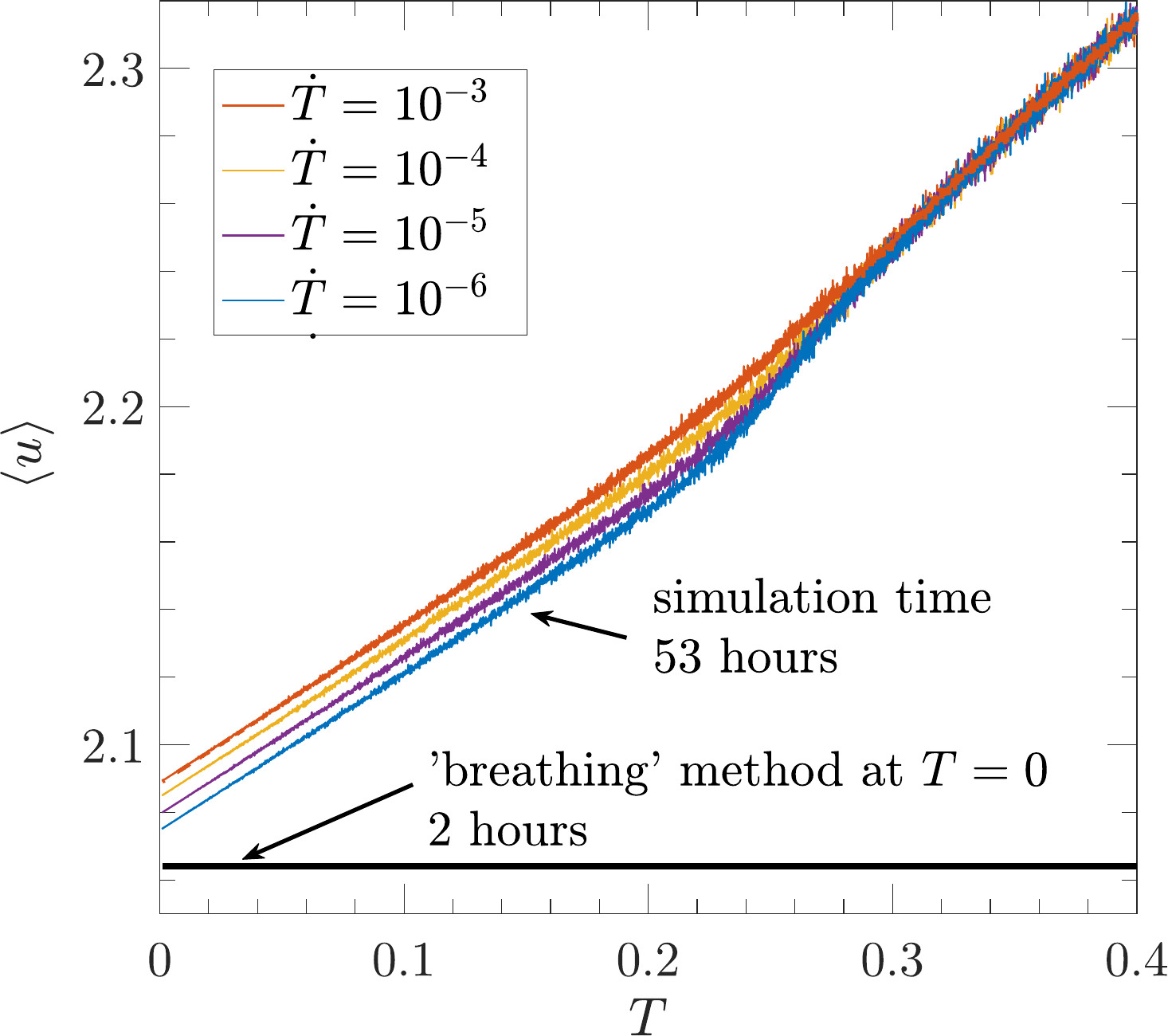}
	\caption{
		Mean interaction potential energy as obtained by sample preparation
		using `breathing' dynamics
		(at $T = 0$ for $K = 10^2$, $t_p = 10^4$, and
		$T_p = 0.2$, in black)
		and using normal dynamics at different cooling rates (cooling from $T=0.4$)
		as indicated in the legend.
		Note that in both cases the ensemble comprises $n = 10$ samples,
		but that normal dynamics are run using smaller than usual
		samples comprising $N = 2000$ particles
		($N = 8000$ is used throughout).
		We verify the representativeness of these smaller samples using
		$N = 8000$ for $\dot{T} = 10^{-3}$, shown using a dashed red line
		(that indeed coincides with the solid red line for $N = 2000$).
		The required CPU time to run the entire simulation with $N = 2000$ particles is indicated. From bottom to top, $\dot{T}$ increases.
		Note that for our `breathing' dynamics the time has been divided by four to correct for
		the difference in system size.
	}
	\label{fig:swap_efficiency}
\end{figure}

We choose $T_p$ and $t_p$ to empirically generate a configuration
in the lowest possible energetic state in terms of
the mean interaction energy
$\langle u \rangle = \langle U \rangle / N$
(averaged on an ensemble of $n = 10$ samples).
In particular, we set $t_p = 10^4$
(the highest value we can practically reach,
with each sample taking eight CPU hours to prepare).
We manually optimise $T_p$ as reported in \cref{fig:Tp}.
Note that we verify that the $T_p$ at which we find the optimum,
is robust in terms of preparation duration $t_p$,
by comparing our results to those for
$t_p = 500$ (dashed line in \cref{fig:Tp}).
Furthermore, the reader is reminded that although
the particle size distribution
depends on temperature while still at $T_p$,
the final
particle size distribution at zero temperature is
independent of $T_p$.
Note that our `breathing' dynamics (at small $K$) are quite efficient to prepare samples
in a low potential energy state.
We verify this by preparing an ensemble
(again $n = 10$, but with $N = 2000$ particles)
with normal dynamics and a slow quench rate.
We plot the potential energy $\langle u \rangle$
at different temperatures in \cref{fig:swap_efficiency}.
In all cases $\langle u \rangle$ at $T = 0$ is higher
than that for the sample prepared using `breathing' dynamics,
which was prepared at a fraction of the computational costs
(sample preparation is a factor of 2000 faster using `breathing' dynamics).

\section{Measurement of quasi-localised modes}
\label{sec:qlm}

\subsection*{Spectrum of the Hessian}
\label{sec:qlm:hessian}

We extract the Hessian (or stiffness matrix) -- the
second derivative of interaction energy -- as follows
\begin{equation}
\mathcal{H}_{ij}
\equiv
\frac{\partial^2 U}{\partial \vec{r}_i \, \partial \vec{r}_j}
=
- \frac{d^2 \varphi(r_{ij})}{d r_{ij}^2}
\frac{\vec{r}_{ij} \, \vec{r}_{ij}}{r_{ij}^2}
- \frac{d \varphi(r_{ij})}{d r_{ij}}
\frac{1}{r_{ij}}
\left(
\mathcal{I} -
\frac{\vec{r}_{ij} \, \vec{r}_{ij}}{r_{ij}^2}
\right)
\end{equation}
for $i \neq j$.
The diagonal
\begin{equation}
\mathcal{H}_{ii} = - \sum_{i \neq j} \mathcal{H}_{ij}
\end{equation}
due to translation symmetry. Note that $\mathcal{H}_{ij}$ is a second-order tensor,
and that $\mathcal{I} = \delta_{\alpha \beta} \vec{e}_\alpha \vec{e}_\beta$
is a second order unit tensor.
We then diagonalise the Hessian, leading to $N$ eigenvalues
$\lambda$ and corresponding eigenmodes $\vec{\Phi}_i$.
Because all particles have a mass $m = 1$
the corresponding $N$ eigenfrequencies are
\begin{equation}
\omega \equiv \sqrt{\lambda}
\end{equation}

We finally represent the spectrum of the Hessian as
\begin{equation}
\label{eq:D(w)}
D(\omega) = \frac{1}{3N - 3}
\sum_{k = 1}^{3N-3} \delta(\omega - \omega_k)
\end{equation}

\subsection*{Density of quasi-localised modes}

The density of quasi-localised modes, $D_L(\omega)$,
follows from the spectrum of the Hessian in \cref{eq:D(w)}
by filtering plane waves that have a frequency
$\omega_e < \omega_c$ (where $\omega_c$ is defined below).
We identify these plane waves by their signature
in participation ratio
\begin{equation}
P_r (\vec{\Phi}_i) \equiv
\frac{(\sum_i || \vec{\Phi}_i ||^2)^2}{N\sum_i || \vec{\Phi}_i ||^4}
\end{equation}
Plane waves thereby have $P_r \approx 2 / 3$, while quasi-localised
modes have $P_r \ll 1$.

In practice, most of our samples have no plane waves below $\omega_c$,
rendering filtering obsolete.
In fact, we only apply filtering after
sample preparation for $K = \{ 10^2, 10^3\}$.
Since we empirically observe the plane waves to be
well separated from the quasi-localised modes in
terms of frequency, we remove them by removing the first
$3 + 12$ eigenmodes of each realisation for $K = 10^3$ and
$3 + 12 + 24$ eigenmodes of each realisation for $K = 10^2$,
corresponding the $3$ translational modes and the first (two)
bands of plane waves\footnote{%
	For $K = 10^3$, $\omega_e = 1.30 \pm 0.02$, and for $K = 10^2$,
	$\omega_e = \{ 1.26 \pm 0.03, 1.73 \pm 0.05 \}$
	(where the uncertainty refers to the standard deviation).
}.
Note that $D_L(\omega)$ is not renormalised after
filtering of plane waves.

We emphasise that in all other measurements
$D_L (\omega) = D(\omega)$ at low frequency.

\subsection*{Protocol to measure \texorpdfstring{$\omega_c$}{omega\_c}}
\label{sec:qlm:omega_c}

\begin{figure}[ht]
	\centering
	\includegraphics[width=0.7\linewidth]{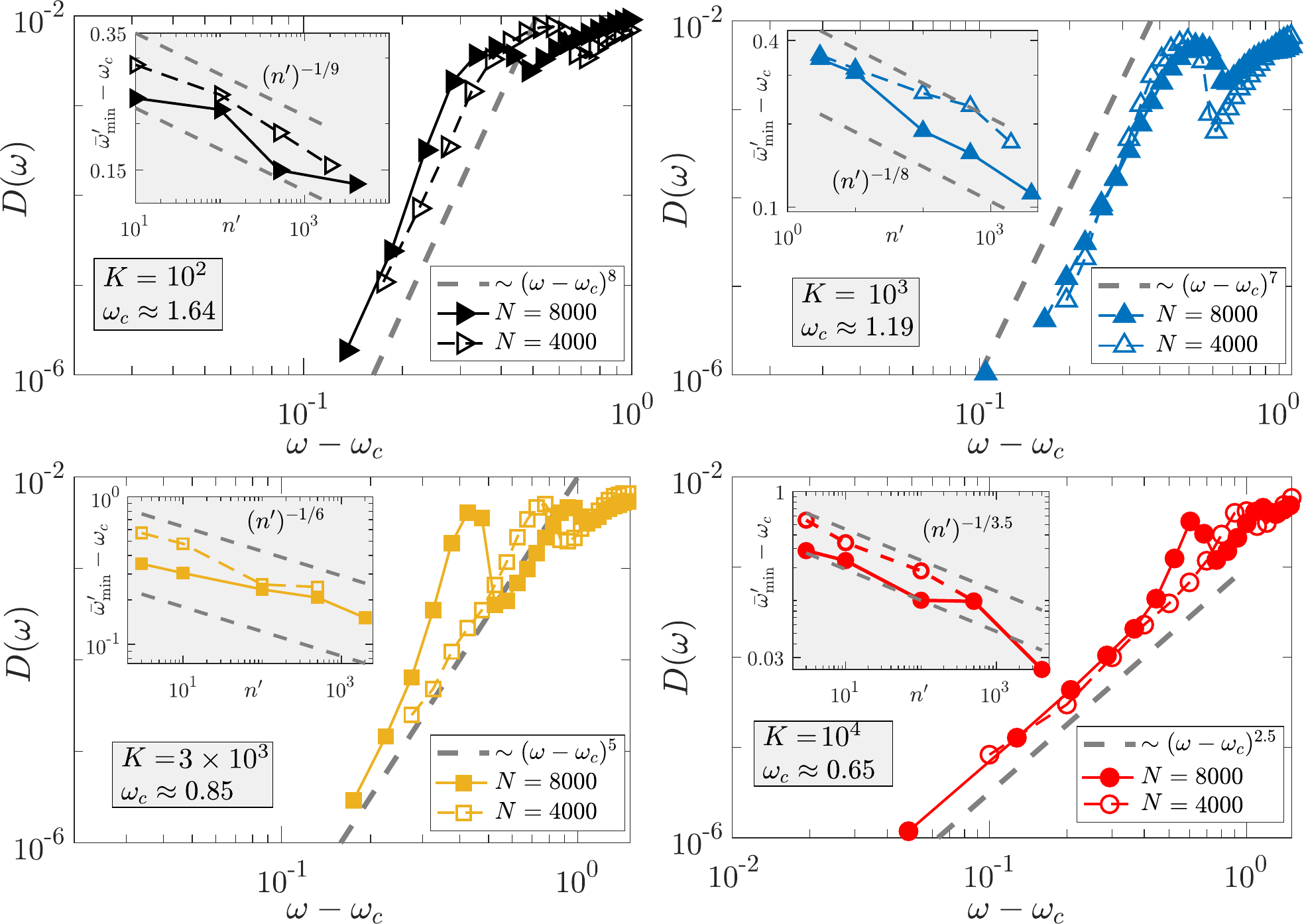}
	\caption{
		Fit of $\omega_c$ by asserting the power law scaling in \cref{eq:omega_c_fit} and
		check by extreme value statistics as in \cref{eq:omega_c_extreme} (insets), for
		all considered $K$.
	}
	\label{fig:fit:omega_c}
\end{figure}

We measure the gap frequency $\omega_c$ -- the frequency of the first quasi-localised mode.
To measure $\omega_c$, we assert that the density
of soft quasi-localised modes follows
\begin{equation}
\label{eq:omega_c_fit}
D_L(\omega) \sim
\left( \omega - \omega_c \right)^{\zeta}
\end{equation}
at low frequency $\omega$.
We then move $\omega_c$
until the power law is most obvious
at low $\omega$, as shown in \cref{fig:fit:omega_c}.
We then visually extract the power $\zeta$ and check
that it and the extracted $\omega_c$ are consistent with extreme-value statistics.
In particular, we expect
\begin{equation}
\label{eq:omega_c_extreme}
\omega_\text{min}' -
\omega_c \sim (n')^{-1 / (1 + \zeta)}
\end{equation}
where $\omega_\text{min}'$ is the frequency of the
softest quasi-localised mode in an ensemble of $n'$ realisations
chosen as a random subset of our ensemble of $n$ realisations.
We consider $\bar{\omega}_\text{min}'$ the average of lowest three
realisations (out of $n'$ realisations).
Indeed, our extracted $\omega_c$ is consistent with this scaling, as shown
in the insets of \cref{fig:fit:omega_c}.
In addition, we check that $\omega_c$ is robust to a change of system size
(shown in \cref{fig:fit:omega_c}).

\subsection*{Protocol to fit \texorpdfstring{$A_4$}{A\_4}}

$A_4$ is extracted from $D_L(\omega)$ by fitting
\begin{equation}
D_L(\omega) = A_4 \, \omega^4
\end{equation}
(\textit{i.e.}~\cref{eq:Dw}) for frequencies below the first plane wave
(for $K = 10^2$) and for frequencies below $\omega_c$
(for $K = 10^3$, $3 \times 10^3$, and $10^4$).
Note that consequently $D_L(\omega) = D(\omega)$
in the relevant frequency range for all these measurements.
The mean and the error of $A_4$
follow as the mean and standard deviation of
$\left\{ \ln D(\omega_i) - 4 \ln\omega_i \right\}$
where $\omega_i$ corresponds to the
position of the bins of $D(\omega)$.

\begin{figure}[ht]
	\centering
	\includegraphics[width=0.7\linewidth]{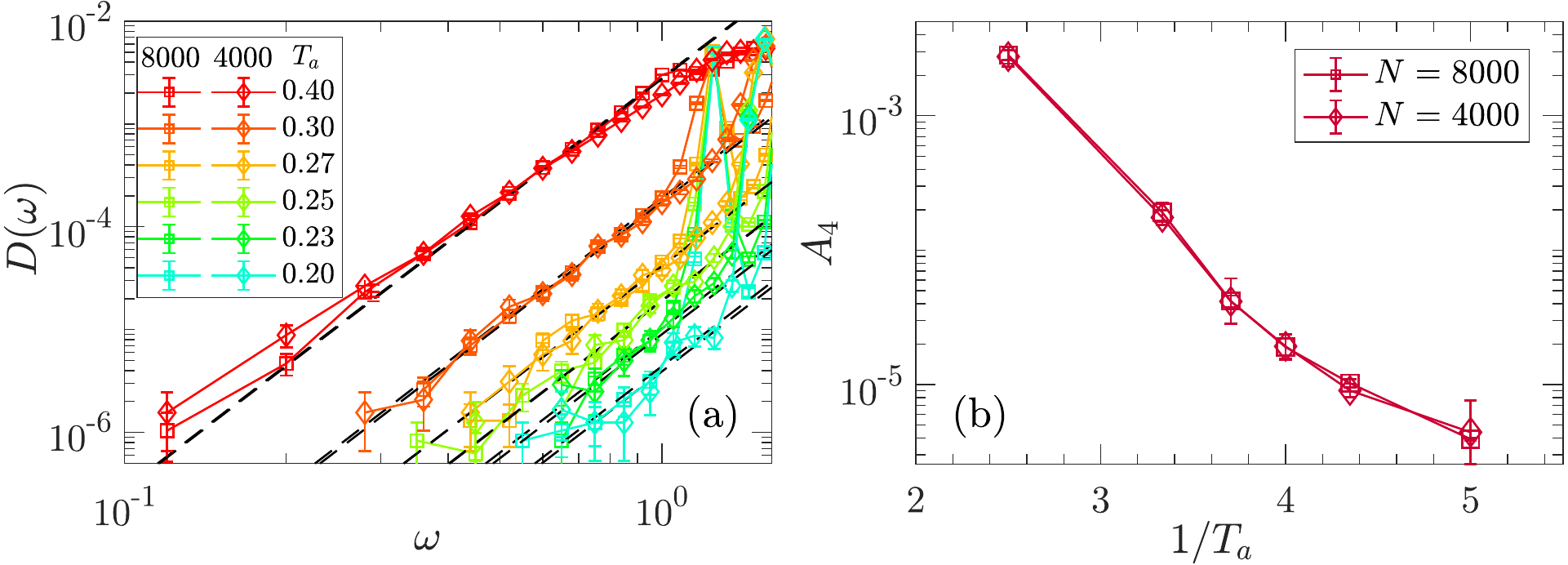}
	\caption{
		(a) $D(\omega)$ for different system sizes $N$ and different activation temperatures $T_a$
		(as indicated in the legend $T_a$ increases from bottom to top).
		(b) $A_4$ fitted on (a) as a function of $1 / T_a$.
		Both plots are for $K = 10^2$ and $t_a = 500$.
	}
	\label{fig:qlm:N}
\end{figure}

We verify that the value of $A_4$ that we fit
is robust to a mild decrease of system size
(using $n = 2000$ realisations of $N = 4000$ particles,
compared to an ensemble of
$n = 4000$ realisations of $N = 8000$ particles).
We find that both the density of soft
quasi-localised modes and the extracted $A_4$
are robust to the change of system size,
as reported in \cref{fig:qlm:N}.

\subsection*{Protocol to fit \texorpdfstring{$E_a$}{E\_a} and \texorpdfstring{$\gamma$}{gamma}}

Our protocol to fit $E_a$ and extract $\gamma$ consists of two steps.
1) We first collapse the curves of $A_4 (T_a)$ for different $t_a$.
Thereto we shift the horizontal axis of e.g.\ \cref{fig:reheat}(b) in
accordance with assuming a functional dependence
\begin{equation}
A_4 = A_4 \left( t_a^\gamma e^{-E_a/T_a} \right)
\end{equation}
until the curves for different $t_a$ collapse to a single
curve (e.g.\ \cref{fig:reheat}(c)), by optimising the ratio
$\gamma / E_a$.
2) On the master curve we next fit $E_a$ of low $T_a$.
Since we know the ratio $\gamma / E_a$, the fitted value of
$E_a$ gives us direct access to $E_a$.
Specifically,
we fit $\ln (A_4)$ \emph{vs} $1 / T_a - \gamma / E_a \ln t_a$
using linear regression to get $E_a$ and its error at low
$T_a$ (the lowest 5 data points in \cref{fig:reheat}(c)).

\subsection*{Results for different \texorpdfstring{$K$}{K}}
\label{sec:qlm:K}

\begin{figure}[ht]
	\centering
	\includegraphics[width=0.9\linewidth]{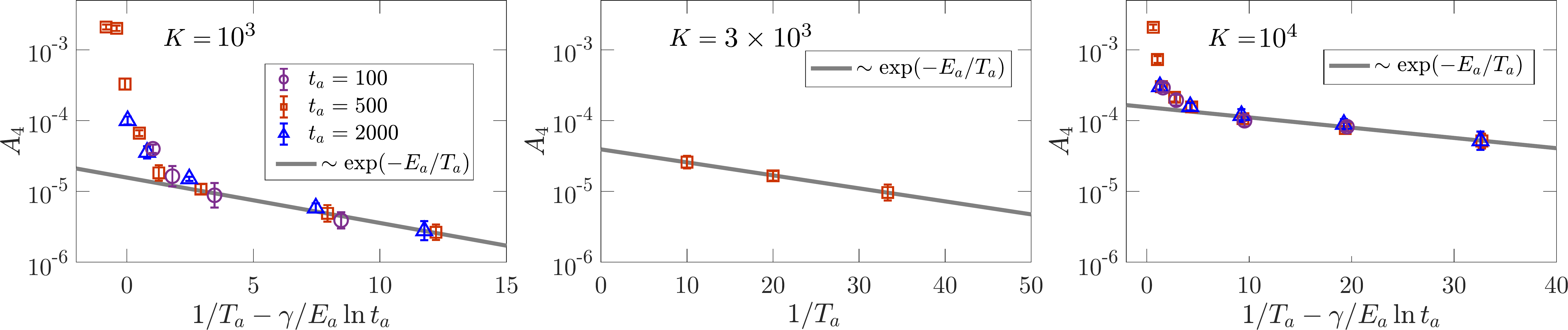}
	\caption{
		Fitting of $E_a$ and $\gamma$ for all $K$ not shown in the main text.
		The fitted values are reported in \cref{sec:nomenclature}.
	}
	\label{fig:qlm:K}
\end{figure}

In \cref{fig:qlm:K} we show the collapse of different waiting times $t_a$
and the fit of $E_a$ at low $T_a$
for all ensembles that are not shown in the main text (notably \cref{fig:reheat}).
Note that for $K = 3 \times 10^3$ we extract $E_a$ by directly fitting
for low $T_a$ for a single $t_a$.

\subsection*{Robustness of \texorpdfstring{$E_a$}{E\_a}}

\begin{figure}[ht]
	\centering
	\includegraphics[width=0.37\linewidth]{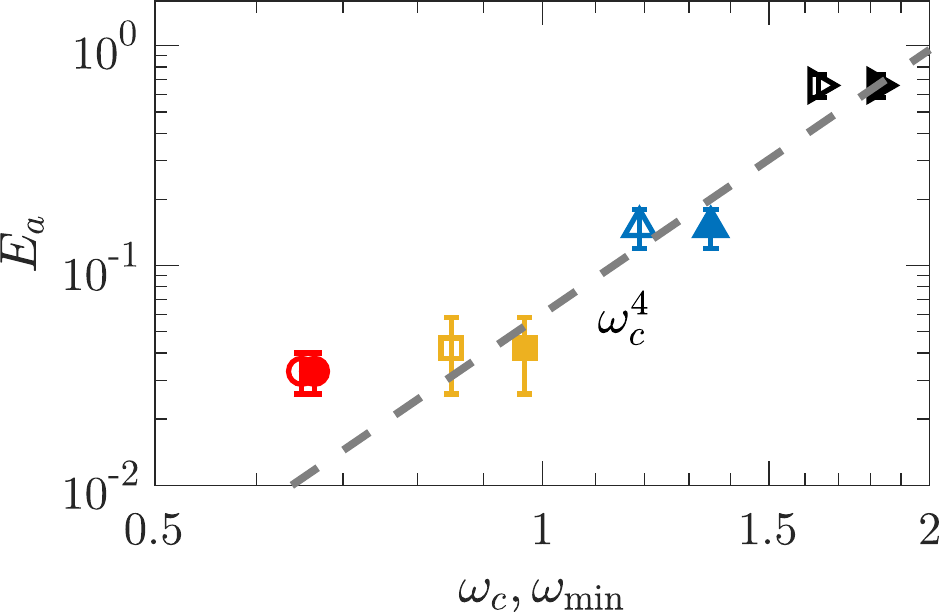}
	\caption{
		$E_a$ as a function of $\omega_c$ (open markers)
		or as a function of $\omega_{\min}$ (solid markers).
	}
	\label{fig:Ea:robustness}
\end{figure}

In \cref{fig:Ea:robustness} we verify that the consistency
with $E_a \sim \omega_c^4$ is robust to a different measure of the
softest quasi-localised mode after sample preparation.
In particular, we compare with $\omega_{\min}$.

\section{The Jacobian of the transformation \texorpdfstring{from $(\lambda_1,\kappa_1,\chi_1)$ to $(\lambda_2,\kappa_2,\chi_2)$}{}}
\label{sec:Jacobian}

\begin{figure}[htp]
	\centering
	\includegraphics[width=.2\linewidth]{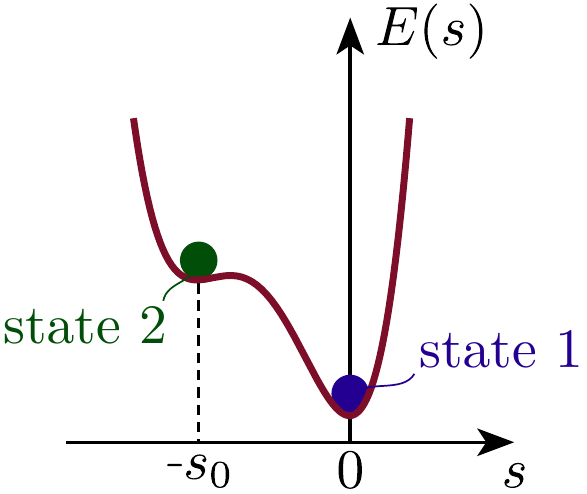}
	\caption{Double-well potential.}
	\label{fig:dwp}
\end{figure}

The potential around the state $1$ is given by:
\begin{equation}
\label{eq:mim_1}
E(s)
= \frac{1}{2!} \lambda_1 s^2
+ \frac{1}{3!} \kappa_1 s^3
+ \frac{1}{4!} \chi_1 s^4
+ \textrm{constant}
\end{equation}
with the joint distribution $P(\lambda_1, \kappa_1, \chi_1)$ that is strictly zero at $\lambda_1 < \omega_c^2$ and smooth above $\omega_c^2$.  $\chi_1>0$. Without loss of generality, we let $\kappa_1>0$.
For the new minimum, state $2$, the potential reads:
\begin{equation}
\label{eq:mim_2}
E(s)
= \frac{1}{2!} \lambda_2 \left(s + s_0 \right)^2
+ \frac{1}{3!} \kappa_2 \left(s + s_0\right)^3
+ \frac{1}{4!} \chi_2 \left(s + s_0\right)^4
+ \textrm{constant}
\end{equation}
where $s_0$ is the shift along $s$, see \cref{fig:dwp}; and
the corresponding joint distribution is $P(\lambda_2, \kappa_2, \chi_2)$. The convention $\kappa_1>0$ leads to the position $s_2=-s_0$ of state $2$  smaller than $0$.

The relation between two sets of coefficients $\lambda_1, \kappa_1, \chi_1$ and
$\lambda_2, \kappa_2, \chi_2$ is
\begin{equation}
\begin{cases}
\label{eq:Jacobian:c}
\chi_1 = \chi_2 \\
\kappa_1 = \kappa_2 + \chi_2 s_0 \\
\lambda_1 = \lambda_2 + \kappa_2 s_0 + \frac{1}{2} \chi_2 s_0^2
\end{cases}
\end{equation}
where $s_0$ as a function of
$(\lambda_2, \kappa_2, \chi_2)$
follows from the fact that the linear term vanishes in \cref{eq:mim_1,eq:mim_2}.
In particular,
\begin{equation}
\label{eq:Jacobian:a}
6 \lambda_2 + 3 \kappa_2 s_0 + \chi_2 s_0^2 = 0
\end{equation}

The joint distribution $P(\lambda_2, \kappa_2, \chi_2)$ is given by
\begin{equation}
P(\lambda_2, \kappa_2, \chi_2) =
\left|
\det
\left(
\frac{d\lambda_1 d\kappa_1 d\chi_1}
{d\lambda_2 d\kappa_2 d\chi_2}
\right)
\right|
P(\lambda_1, \kappa_1, \chi_1)
\end{equation}
where
\begin{equation}
\left|
\det \left(
\frac{d\lambda_1 d\kappa_1d\chi_1}
{d\lambda_2 d\kappa_2d\chi_2}
\right)
\right|
\equiv
\left|\det
\left(
\begin{bmatrix}
\frac{\partial\lambda_1}{\partial\lambda_2} &
\frac{\partial\lambda_1}{\partial\kappa_2} &
\frac{\partial\lambda_1}{\partial\chi_2}
\\
\frac{\partial\kappa_1}{\partial\lambda_2} &
\frac{\partial\kappa_1}{\partial\kappa_2} &
\frac{\partial\kappa_1}{\partial\chi_2}
\\
\frac{\partial\chi_1}{\partial\lambda_2} &
\frac{\partial\chi_1}{\partial\kappa_2} &
\frac{\partial\chi_1}{\partial\chi_2}
\end{bmatrix}
\right)
\right|
=
\left|
1 + \kappa_2
\frac{\partial s_0}{\partial\lambda_2}
+ \chi_2\frac{\partial s_0}{\partial\kappa_2}
\right|
\end{equation}

From \cref{eq:Jacobian:a} we find that
\begin{align}
\frac{\partial s_0}{\partial\lambda_2}=-\frac{6}{3\kappa_2+2\chi_2 s_0}
\\
\frac{\partial s_0}{\partial\kappa_2}=-\frac{3s_0}{3\kappa_2+2\chi_2 s_0}
\end{align}

And thus:
\begin{equation}
\label{eq:Jacobian:b}
\left|
\det
\left(
\frac{d\lambda_1 d\kappa_1d\chi_1}
{d\lambda_2 d\kappa_2d\chi_2}
\right)
\right|
=
\left|
\frac{-3 \kappa_2 - \chi_2 s_0}
{3 \kappa_2 + 2 \chi_2 s_0}
\right|
=
\left|
\frac{6 \lambda_2}
{3 \kappa_2 s_0 + 2 \chi_2 s_0^2}
\right|
\end{equation}

If the excited state of the double-well potential is close to the spinodal case,
$\lambda_2$ is small (as it is in \cref{fig:dwp}).
In particular, when $\lambda_2 \approx 0$, it follows that
$s_0 \approx -3 \kappa_2 / \chi_2$.
Inserting this in \cref{eq:Jacobian:b} gives:
\begin{equation}
\left|
\det
\left(
\frac{d\lambda_1 d\kappa_1d\chi_1}
{d\lambda_2 d\kappa_2d\chi_2}
\right)
\right|
\simeq
\frac{2 \lambda_2 \chi_2}
{3 \kappa_2^2}
\simeq \frac{\lambda_2}{\lambda_1}
\sim
\lambda_2
\end{equation}
where $\lambda_1$ is a large value with the lower bound $\omega_c^2$.
We have thus found that the joint distribution
\begin{equation}
P(\lambda_2, \kappa_2, \chi_2)
\sim \lambda_2 \,
P(\lambda_1, \kappa_1, \chi_1)
\end{equation}

Hence, the marginal distribution $P(\lambda_2)\sim \lambda_2$ (after
integrating out $\kappa_2$ and $\chi_2$) and therefore
\begin{equation}
D(\omega_2)
= P(\lambda_2) \, \frac{d\lambda_2}{d\omega_2}
\sim \omega_2^3 .
\end{equation}

\subsection*{Gap in energy barrier distribution}

For a given $\lambda_1$ we define $c(\kappa_1,\chi_1)=\lambda_1\chi_1/\kappa_1^2$,
which smoothly varies in a narrow range from $1/3$ (for a symmetric double-well) to $3/8$
(for a spinodal).
Then we can express the energy barrier as
\begin{equation}
\Delta E = \lambda_1^2
\frac{
	\left(3 - \sqrt{9 - 24 c}\right)^2 \left(-3 + 12 c + \sqrt{9 - 24 c}\right)
}{
	192 \, \chi_1 \, c^2
}
\end{equation}
The function $c$ is slowly varying and not singular so that
\begin{align}
\Delta E \sim \lambda_1^2\sim \omega_c^4
\end{align}
Similarly, the energy difference
\begin{equation}
E_{12} = \lambda_1^2
\frac{
	\left(3 + \sqrt{9 - 24 c}\right)^2 \left(-3 + 12 c - \sqrt{9 - 24 c}\right)
}{
	192 \, \chi_1 \, c^2
}
\sim \lambda_1^2 \sim \omega_c^4
\end{equation}
(except in the case of a symmetric double-well).

\section{Estimation of \texorpdfstring{$\chi_1$}{chi}}
\label{sec:estimation}

Numerically, the coefficient $\chi_1$ (from \cref{eq:Taylor}) along the direction of the displacement field $\vec{s}$ from the initial minimum to the new minimum whose frequency is smaller than $\omega_e$ (the frequency of first tranverse plane waves), can be expressed by the pair interaction $\varphi(\vec{r})$ at mechanical equilibrium, as follows:
\begin{align}
\chi_1
=& \sum_{\alpha,\beta,\eta,\nu = 1}^3 \; \sum_{n,m,k,l=1}^N \frac{\partial^{4}U}{\partial r_{n}^{\nu}\partial r_{m}^{\eta}\partial r_{k}^{\beta}\partial r_{l}^{\alpha}}s_{l}^{\alpha}s_{k}^{\beta}s_{m}^{\eta}s_{n}^{\nu}\\
= & \sum_{i<j} \Bigg\{
\left(
\frac{1}{r_{ij}^4} \frac{d^4 \varphi}{d r^4}
- \frac{6}{r_{ij}^5} \frac{d^3 \varphi}{d r^3}
+ \frac{15}{r_{ij}^6} \frac{d^2 \varphi}{d r^2}
- \frac{15}{r_{ij}^7} \varphi'
\right)
\big( \vec{r}_{ij} \cdot \vec{s}_{ij} \big)^4
\\
& +
\left(
\frac{1}{r_{ij}^3} \frac{d^3 \varphi}{d r^3}
- \frac{3}{r_{ij}^4} \frac{d^2 \varphi}{d r^2}
+ \frac{3}{r_{ij}^5} \varphi'
\right)
6 \, \big( \vec{r}_{ij} \cdot \vec{s}_{ij} \big)^{2}
\big( \vec{s}_{ij} \cdot \vec{s}_{ij} \big)
+
\left(
\frac{1}{r_{ij}^2} \frac{d^2 \varphi}{d r^2}
- \frac{1}{r_{ij}^3} \varphi'
\right)
3 \, \big( \vec{s}_{ij} \cdot \vec{s}_{ij} \big)^{2}
\Bigg\}
\end{align}
where $\vec{r}_{i}$ are the particles' equilibrium positions, $\vec{s}$ is the direction (the normalised displacement field from the quenched states before and after reheating), and $\vec{s}_{ij}=\vec{s}_{i}-\vec{s}_{j}$.

\begin{figure}[ht]
	\centering
	\includegraphics[width=0.4\linewidth]{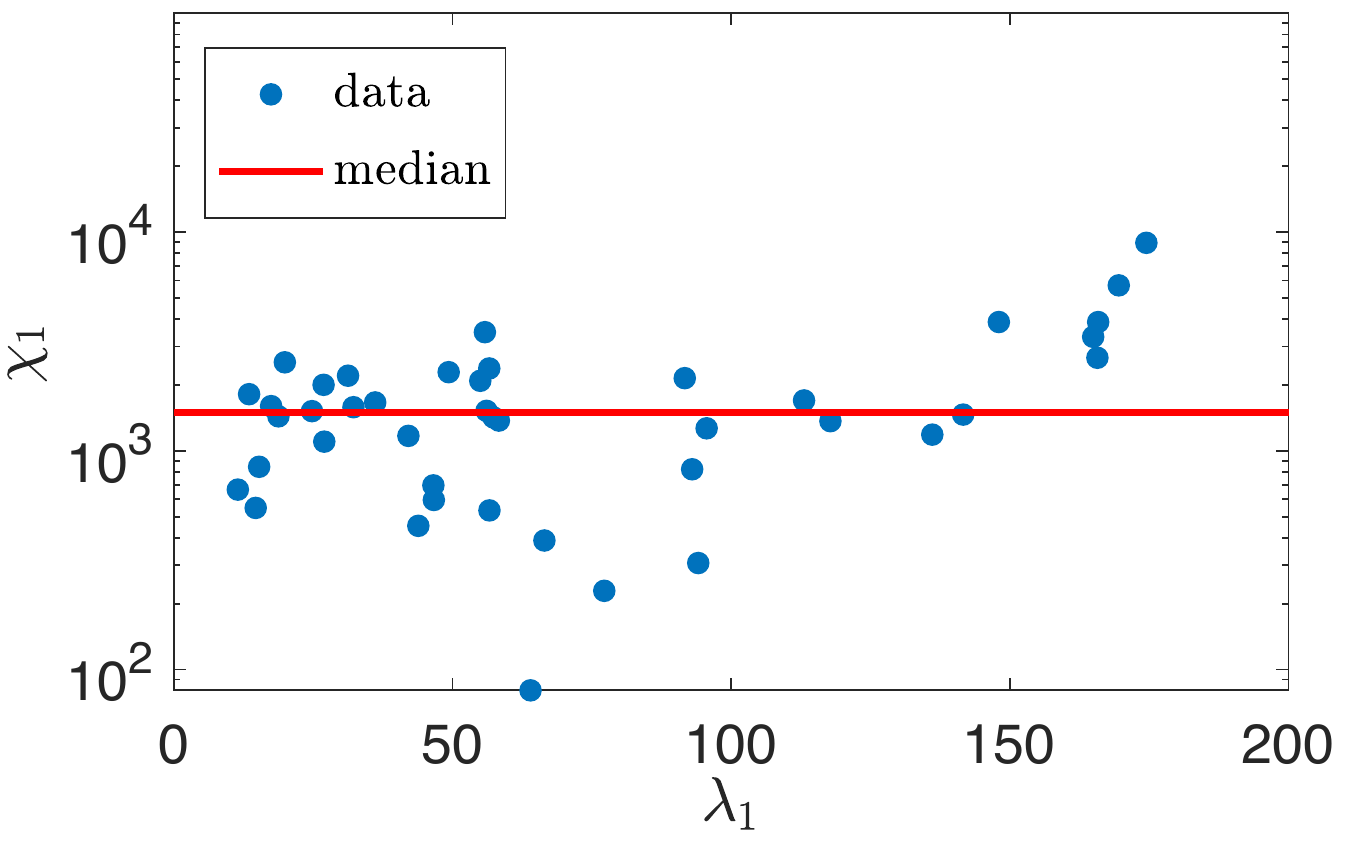}
	\caption{Scatter plot $\chi_1$ as function of $\lambda_1$ for our largest gap ($\omega_c = 1.64$). $\lambda_1$ follows as $\lambda_{1}=\sum_{\alpha,\beta=1}^3 \sum_{k,l=1}^N s_{l}^{\alpha}H_{lk}^{\alpha\beta}s_{k}^{\beta}$.}
	\label{fig:chi}
\end{figure}

From the results in \cref{fig:chi} we observe that at small $\lambda_1$, $\chi_1$ is independent of $\lambda_1$. Hence, we regard it as a constant. Here we estimate $\chi_1 \approx 1500$ by its median value. Note that $\chi_1$ has unit $m \omega_0^2 / d_0^2$, where $m$ is the particle mass, $d_0$ is approximately equal to the inter-particle distance $a$ (hence we take $a = d_0$), and $\omega_0\equiv1/t_0$ is the unit frequency in our simulation which is about $\omega_D/18$. Here $\omega_D$ is the Debye frequency.
Hence we get $\chi_1\approx 1500 m \omega_0^2/a^2\approx 4.6 m\omega_D^2/a^2$.

The relation between $\omega_D$ and $\omega_0$ is calculated by: $\omega_{D} = \left(9N / (4\pi(2\omega_{e}^{-3}+\omega_{l}^{-3})) \right)^{1/3}\approx18\omega_0$,  where $\omega_e=1.26 \omega_0$ and $\omega_l\approx 2.5\omega_e$ ($\omega_l$ is the frequency of first longitudinal modes). Note that both are plane waves. In particular, the first transverse modes consist of 12 modes and the first longitudinal modes consist 6 modes. Their identification is straightforward through the number of modes and their participation ratios, which are around 0.6. Note that $\omega_0=1$ in our simulation, so $\omega_D \approx 18$ and $\omega_c / \omega_D \approx 0.1$.

\section{Geometry of rearrangements}
\label{sec:geometry}

\subsection*{Protocol to separate rearrangements}

The displacement field between the states before and after reheating
may contain more than one elementary excitation.
We extract them one-by-one from this displacement field, by assuming them
linearly independent.
This corresponds to the following algorithm:

\begin{enumerate}
	\item
	Find the particle with the largest displacement.

	\item
	Place a small sphere centred at this particle with a radius
	$\tilde{R}^{(i)} = (V/N)^{1/3}$
	(with $i$ the increment number, starting at $i = 0$).

	\item
	Set all displacements outside the sphere equal to zero.
	The particle displacements inside the sphere are not changed.

	\item
	Minimise the energy $U$ (every particle is free to move).

	\item
	Increase the radius of sphere:
	$\tilde{R}^{(i + 1)} = \tilde{R}^{(i)} + \Delta \tilde{R}$,
	and reset the displacements as in step 3
	(the particle displacements outside the sphere are
	set to zero and those inside the sphere equal to the
	original particle displacements).

	\item
	Repeat steps 4 and 5, until the localised mode is identified.
	In particular, stop when
	$|U^{(i+1)} - U^{(i)}| < 10^{-6}$ and,
	to avoid stopping too early,
	the norm of the displacement field is larger than $10^{-2}$.
	Note that $U^{(i)}$ refers to the potential energy after energy
	minimisation, in step 4, for increment $i$.

\end{enumerate}

The local rearrangement is then the displacement field
after the last energy minimisation.
We then subtract it from the original displacement
and continue to extract the next elementary excitation,
by repeating this algorithm.
We continue to do so until we have extracted all elementary excitations.
In particular, we stop $\tilde{R}^{(i)} / L > \sqrt{3}/2$
(with $L$ the linear size of the simulation box).

\subsection*{Results}

\begin{figure}[ht]
	\centering
	\includegraphics[width=\linewidth]{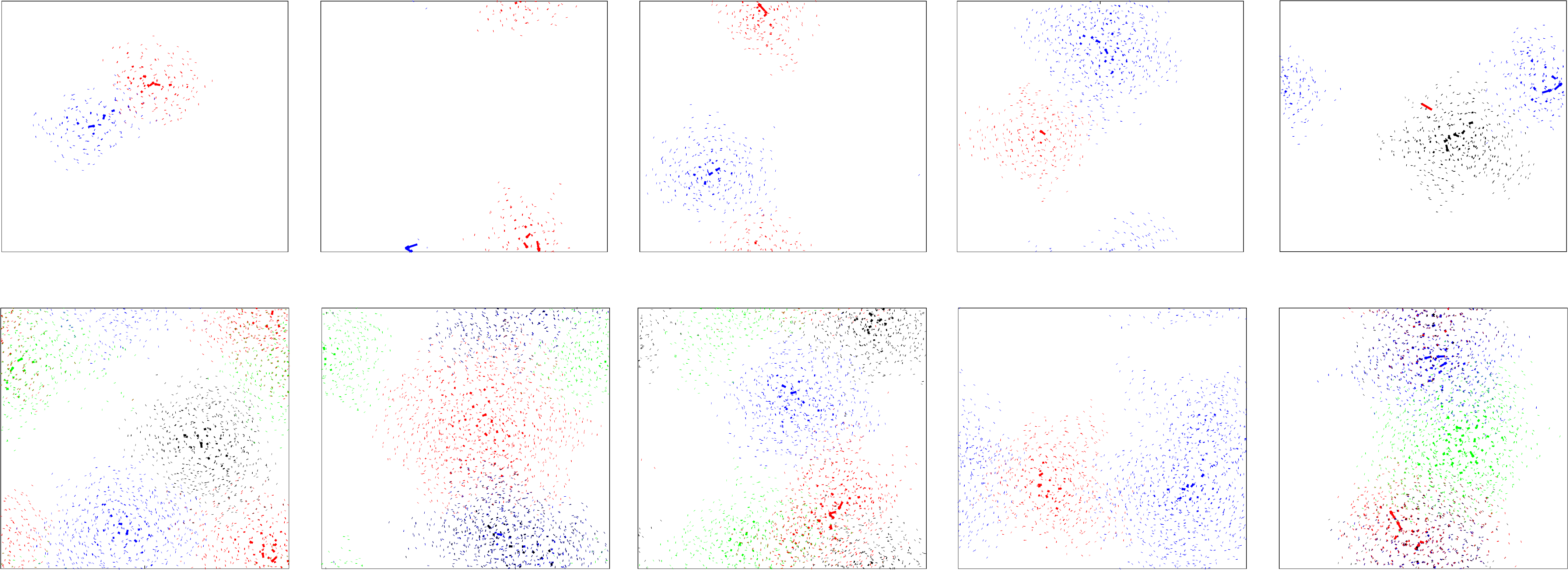}
	\caption{
		Individual local rearrangements projected on the $xy$ plane,
		shown using different colours, in five
		randomly chosen examples
		(from those samples that show more than one local rearrangement)
		for: (top) $K = 10^2$, $T_a = 0.15$, and $t_a =500$,
		(bottom) $K = 10^4$, $T_a = 0.03$, and $t_a =500$.}
	\label{fig:separate}
\end{figure}

Five representative samples (for two different $K$) showing our separation protocol
are shown in \cref{fig:separate},
whereby the displacement field of each elementary excitation is plotted using a different colour.
On average, we measure
$\{ 1.4, 1.6, 1.9, 3.5 \}$ elementary excitations for
$K = \{ 10^2, 10^3, 3 \times 10^3, 10^4\}$ at the lowest thermal activation
$T_a = \{0.15, 0.07, 0.03, 0.03\}$ with $t_a = 500$.

\begin{figure}[ht]
	\centering
	\includegraphics[width=0.65\linewidth]{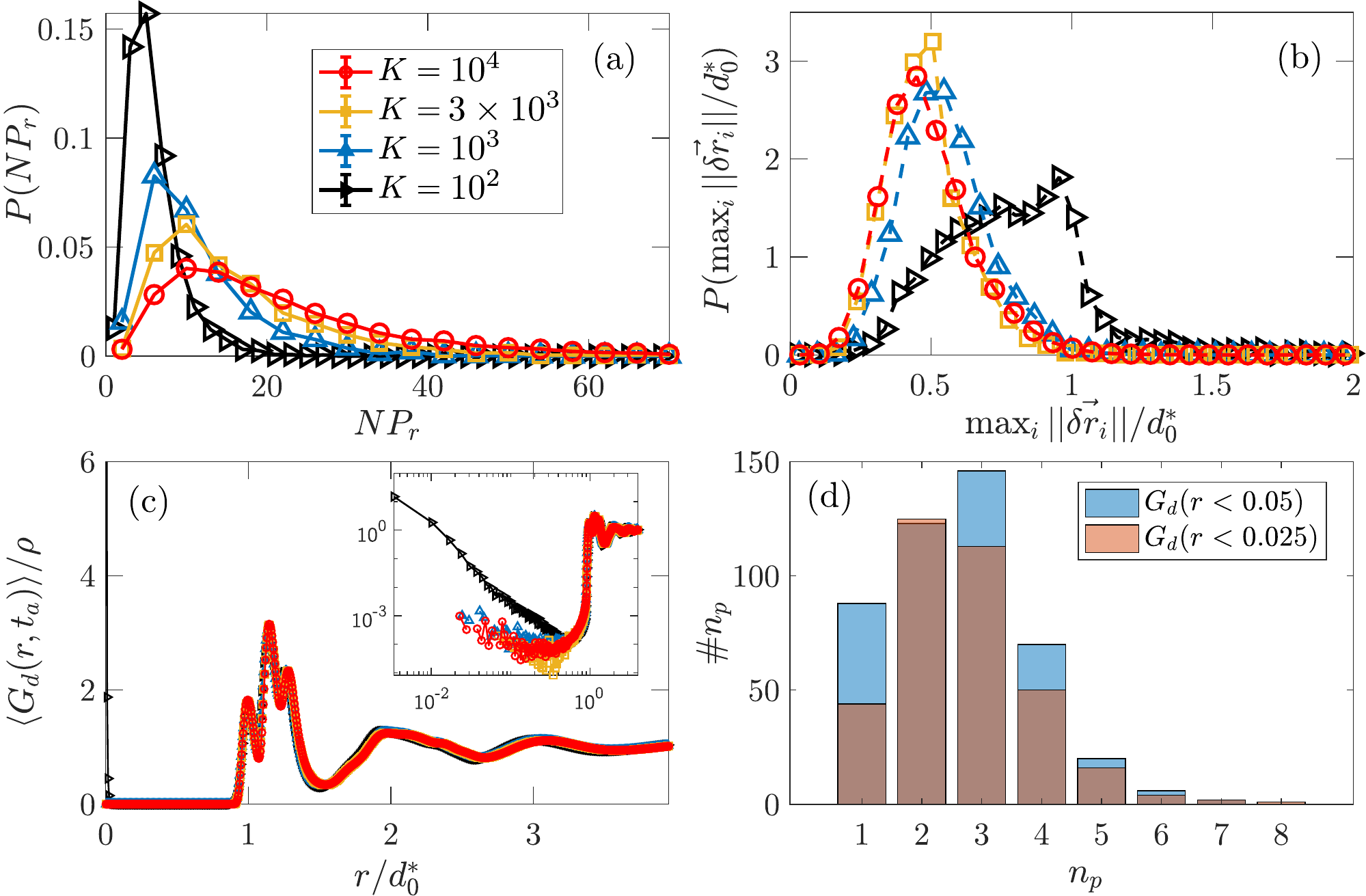}
	\caption{
		Probability distribution of
		(a) the participation ratio $NP_r$ and
		(b) the maximal particle displacement ${\max_i \{|| \delta \vec{r}_i ||\}}$
		at different $\omega_c$.
		(c) The distinct part of the Van Hove correlation normalised by the number density
		$\langle G_d \rangle / \rho $
		at different $\omega_c$.
		(d) Histogram of the number of particles that permute per realisation, $\# n_p$
		(for largest $\omega_c$)
		at two different cutoff distances $r_c / d_0^* = \{0.025, 0.05\}$.}
	\label{fig:distribution}
\end{figure}

We, furthermore, include the distribution of the participation ratio
of the elementary excitations in \cref{fig:distribution}(a),
whereby the different colours correspond to the different data points in \cref{fig:geometry}(a).
We observe that the elementary excitations become more localised for larger gaps.
Likewise, we include the distribution of the maximum displacement of each elementary excitation
in \cref{fig:distribution}(b)
(the different colours correspond to the different data points in \cref{fig:geometry}(b)).
In this case we observe that the maximum displacement increases for our largest $\omega_c$ (in black)
as the result of string-like motion.
This is supported by the distinct part of the Van Hove correlation in \cref{fig:distribution}(c)
that displays a sharp peak around $r = 0$ only for our largest $\omega_c$ (in black).
For this configuration,
we plot the distribution of the number of permuting particles, $\# n_p$, inside the `string'
in \cref{fig:distribution}(d).

\end{document}